\documentclass[10pt]{iopart}

\expandafter\let\csname equation*\endcsname\relax
\expandafter\let\csname endequation*\endcsname\relax
\usepackage{graphicx}
\usepackage{epstopdf}
\usepackage{amssymb}
\usepackage{amsmath}
\usepackage{textcomp}
\usepackage[dvipsnames]{xcolor}
\definecolor{green}{rgb}{0.0, 0.5, 0.0}
\usepackage{booktabs}
\usepackage{cite}
\usepackage{ulem}
\usepackage{tikz}
\usepackage{url}

\begin{document}
\PSST

\title[Fully kinetic model of plasma expansion in a magnetic nozzle]{Fully kinetic model of plasma expansion in a magnetic nozzle}

\author{Shaun Andrews$^{1,2}$, Simone Di Fede$^3$ and Mirko Magarotto$^3$}

\address{$^1$ Department of Industrial Engineering, University of Padova, Via Gradenigo 6/a 35131 Padova, Italy}
\address{$^2$ Technology for Propulsion and Innovation (T4i) S.p.A., Via della Croce Rossa 112 35129 Padova, Italy}
\address{$^3$ Centro di Ateneo di Studi e Attività Spaziali ‘Giuseppe Colombo’ – CISAS, University of Padova, Via Venezia 15 35131 Padova, Italy}

\ead{sa15339@my.bristol.ac.uk}
\vspace{10pt}
\begin{indented}
\item[]November 2021
\end{indented}

\begin{abstract}

A self-consistent model is presented for performing steady-state fully kinetic Particle-in-Cell simulations of magnetised plasma plumes. An energy-based electron reflection prevents the numerical pump instability associated with a typical open-outflow boundary, and is shown to be sufficiently general that both the plume kinetics and plasma potential demonstrate domain independence (within 4\%). This is upheld by non-stationary Robin-type boundary conditions on the Poisson's equation, coupled to a capacitive circuit that allows physical evolution of the downstream potential drop in the transient. The method has been validated against experiments, providing results that fall within the uncertainty of measurements. Simulations are then carried out to study collisional xenon discharges into axisymmetric diverging magnetic nozzles. Particular discussion is given to the identification of a potential well arising from charge separation at the edge of the plume, the role of ion-neutral charge exchange, and a three-region piecewise polytropic cooling regime for electrons. The polytropic index is shown to depend on the degree of magnetisation. Specifically, in the region near the thruster outlet, the plume is weakly-magnetised due to the cross-field diffusion of electron-heavy particle collisions. Downstream, a strongly-magnetised region of near-isothermal expansion occurs. Finally, in the detached region, the polytropic index tends to that of a more adiabatic unmagnetised case. With an increasing magnetic nozzle field strength, an inferior limit is found to the average polytropic index of $\bar{\gamma}_e\sim1.16$.

\end{abstract}

\vspace{2pc}

%\submitto{\PSST}

\vspace{1cm}
\textit{\color{red}This is the version of the article before peer review or editing, as submitted by an author to Plasma Sources Science and Technology. IOP Publishing Ltd is not responsible for any errors or omissions in this version of the manuscript or any version derived from it. The Version of Record is available online at \url{10.1088/1361-6595/ac56ec}.}
\maketitle
\ioptwocol
\section{Introduction}
\label{sec:intro}

The study of electric propulsion continues to receive much attention despite mature technologies such as Ion and Hall effect thrusters establishing dominant flight heritage over the last two decades. However, such systems are increasingly being recognised as complex and high cost, particularly for small-satellite {applications} \cite{b:keidar2014}. Therefore, in the last few years, particular effort has been made in the development of magnetically-enhanced plasma thrusters (MEPT). {This broad category includes the Helicon Plasma Thruster (HPT) \cite{takahashi2019helicon,b:boswell2003,b:shinohara2014,b:merino2015}, the Electron Cyclotron Resonance Thruster (ECRT) \cite{b:cannat2015}, and the Applied Field Magnetoplasmadynamic Thruster (AF-MPDT) \cite{b:boxberger2019}. In such systems the {plasma acceleration} is driven by a magnetic nozzle (MN) \cite{b:merino2016}: a divergent magneto-static field generated by a set of solenoids or permanent magnets. {The MN radially confines the hot partially-magnetised plasma beam and accelerates it supersonically via the conversion of thermal energy into directed axial kinetic energy, therefore enhancing thrust \cite{b:ahedo2010}.}

The HPT and ECRT are cathode-less devices, relying on electromagnetic waves for plasma production and heating \cite{b:chen2015,b:magarotto2019dep}, while the AF-MPDT relies on annular electrodes. Since the resulting plasma beam is quasi-neutral, no additional neutraliser (e.g. a hollow cathode) is required. Thus, MEPTs are becoming an increasing option for low-thrust propulsion, being highly scalable, robust, light, low-cost and resistant to lifetime-limiting erosion \cite{b:manente2019}. MNs also have no physical walls, thus avoiding thermal loading and erosion issues. The first in-orbit demonstration of a radio-frequency MEPT took place from March 2021 by Technology for Innovation and Propulsion (T4i) S.p.A. with the 50~W ‘‘REGULUS’’ thruster  \cite{b:bellomo2021,b:manente2019}. At the same time, the present disadvantage of MEPTs is the relatively low thrust efficiency, generally $<$20\% \cite{takahashi2019helicon}. For MEPTs to be sufficiently efficient and competitive, a high ionisation ratio is mandatory (with electron temperatures of tens of eV \cite{b:magarotto2020,b:magarotto2020_2,b:magarotto2020_3}) or else the specific impulse achievable with the MN is limited.

{The main MN physics is reasonably established and well-understood. In typical MEPTs, with magnetic fields in the 100-1000 G range, ions are weakly magnetised and are bound to the highly magnetised electrons through an ambipolar electric field, which develops to maintain quasi-neutrality \cite{b:longmier2011,b:olsen2015}.} {This results in a potential drop, both radially and axially, which confines most of the electron population while accelerating ions freely downstream. The potential drop self-consistently evolves to maintain a globally current-free plasma, ultimately determining the velocity of ions \cite{b:lafleur2015}}. Nevertheless, there are many other aspects requiring further detailed investigation, such as the evolution of velocity distribution functions (VDF), plasma detachment, anisotropic electron cooling, doubly-trapped electron populations and the role of collisions \cite{b:kaganovich2020}.

Numerical efforts to understand MNs have involved both fluid and kinetic models, some of which make use of semi-analytical solutions \cite{b:merino2018kinetic,b:sasoh1994}. Two-dimensional (2D) fluid models have shown to be a powerful tool to understand the main phenomena \cite{b:ahedo2010}. {However, their closure (i.e. a definition of non-local heat conduction) remains an elusive problem.} One-dimensional (1D) stationary kinetic models of a MN have allowed analysis of the downstream ion and electron heat fluxes and the response to non-Maxwellian features of the ion and electron VDFs \cite{b:martinez2015,b:ahedo2020}. However, except for 1D cases, solving the Boltzmann equation directly is often computationally intensive \cite{b:kim2005}. {Both fluid and kinetic continuum approaches must further make assumptions regarding the VDF, one of the main impact parameters in magnetised plasma expansion \cite{b:ahedo2020}. {Hence}, numerical studies need to be extended to fully kinetic \cite{b:difede2021,b:iac2021,b:porto2019,b:bipic} or fluid-kinetic \cite{b:alvaro2021} approaches if the dynamics of a MN want to be treated self-consistently.} The fully kinetic Particle-in-Cell (PIC) method represents the numerical strategy with the lowest level of assumptions. Both electron and ion populations are modelled as macro-particles, subject to the action of {self-consistently computed} electric and magnetic fields, as well as particle collisions \cite{b:birdsall1991}.

{PIC simulations operate by necessity on a finite domain. Due to the ambipolar potential drop along the MN, and for a typical meso-thermal plume (electron thermal velocity greatly exceeds the ion drift velocity and ion thermal velocity),} {most of the electron population will become trapped into the much-slower ion beam and reciprocate within the plume \cite{b:li2019,b:merino2019iepc}. Since the computational domain is finite, electrons may reverse their trajectories beyond the domain. If these electrons are non-physically deleted upon reaching the open boundaries, the so-called ‘‘numerical pump instability’’ will arise \cite{b:brieda2018,b:briedaThesis}.} For this reason, simulations are usually stopped long before the ion beam reaches the open boundaries. {Thus, most results in the literature deal with short time-scale transient plume expansions in small domains \cite{b:levin2018, b:hu2017}.}

{Alternatively, to prevent the instability, an open model has been demonstrated where a virtual ion sink was implemented midway between the inlet and outer boundaries \cite{b:brieda2018}. Ions are absorbed by the sink, while electrons can permeate through it, thus retaining the trapped electrons between the ion sink and the boundaries. Electrons are also reflected from the boundaries, based on global charge conservation. However, the sink must be located far from these boundaries, resulting in an unwelcome increase in the domain size.} To overcome this limitation, another charge-conserving boundary condition has been proposed \cite{b:levin2020}, where the number of electrons reflected at the open boundary is {determined so as to maintain a globally-neutral plasma}. Another approach, formulated to mimic the real physics, uses a current-free boundary condition \cite{b:li2019}. This has been demonstrated via simulations of both a non-magnetised plume \cite{b:li2019} and a MN \cite{b:chen2020}. {The methodology proposed \cite{b:li2019,b:chen2020} is well-founded concerning electron kinetics, but simplified conditions have been assumed for the solution of the electric field.} A zero-Neumann condition was imposed at the open boundary to solve the Poisson's equation.} Such a condition is appropriate only in the limit of an infinitely large domain that encompasses nearly all the potential fall that occurs in the plasma plume. For this reason, the authors of \cite{b:li2019} suggest ignoring the plasma dynamics in some portion of the domain near the open boundaries (e.g., $\sim$20\%). Indeed, this boundary condition does not generally provide results which are domain independent. {A definition of consistent boundary conditions for treating both magnetised and unmagnetised plumes remains a challenging problem, not yet fully solved.}

{This article presents a new electrostatic fully kinetic PIC model for MN plasma expansions. Boundary conditions are {introduced} to improve on previous works in terms of {both} electron kinetics and the Poisson's equation. Regarding the treatment of the electrons, a consistent approach has been defined to selectively reflect or absorb parts of the population crossing the open boundaries.} {The proposed approach mimics the partial reflection of electrons that would take place further downstream (outside of the domain), by enforcing the integral current-free condition along the open boundaries and an energy-based reflection criterion. The total potential drop is self-consistently calculated to maintain the net-zero current and is included when determining a non-stationary Robin boundary condition on the plasma potential. The result is a set of mutually consistent boundary conditions, which is sufficiently general such that both the plume kinetics, and plasma potential distribution, are independent of the computational domain size and in good agreement with experiments \cite{b:lafleur2011}.}

\Sref{sec:model} summarises the key aspects of the PIC model and introduces the new boundary condition treatment. {In \sref{sec:verif} the capability to produce a stable steady-state plume and a {domain-independent solution is demonstrated}. In \sref{sec:valid} the numerical approach has been benchmarked against measures of plasma density and plasma potential \cite{b:lafleur2011}. In \sref{sec:results} the validated approach has been exploited to investigate the plasma expansion in a MN. The most relevant aspects analysed are: the presence of a collisionally-enhanced radial potential well that {confines the plasma expansion}, the influence of the magnetic field intensity on the propulsive performance, and {the electron cooling.}} The conclusions are then given in \sref{sec:conclude}.
\section{Physical and numerical model}
\label{sec:model}
\begin{figure*}[!tb]
    \centering
    \includegraphics[width=\linewidth]{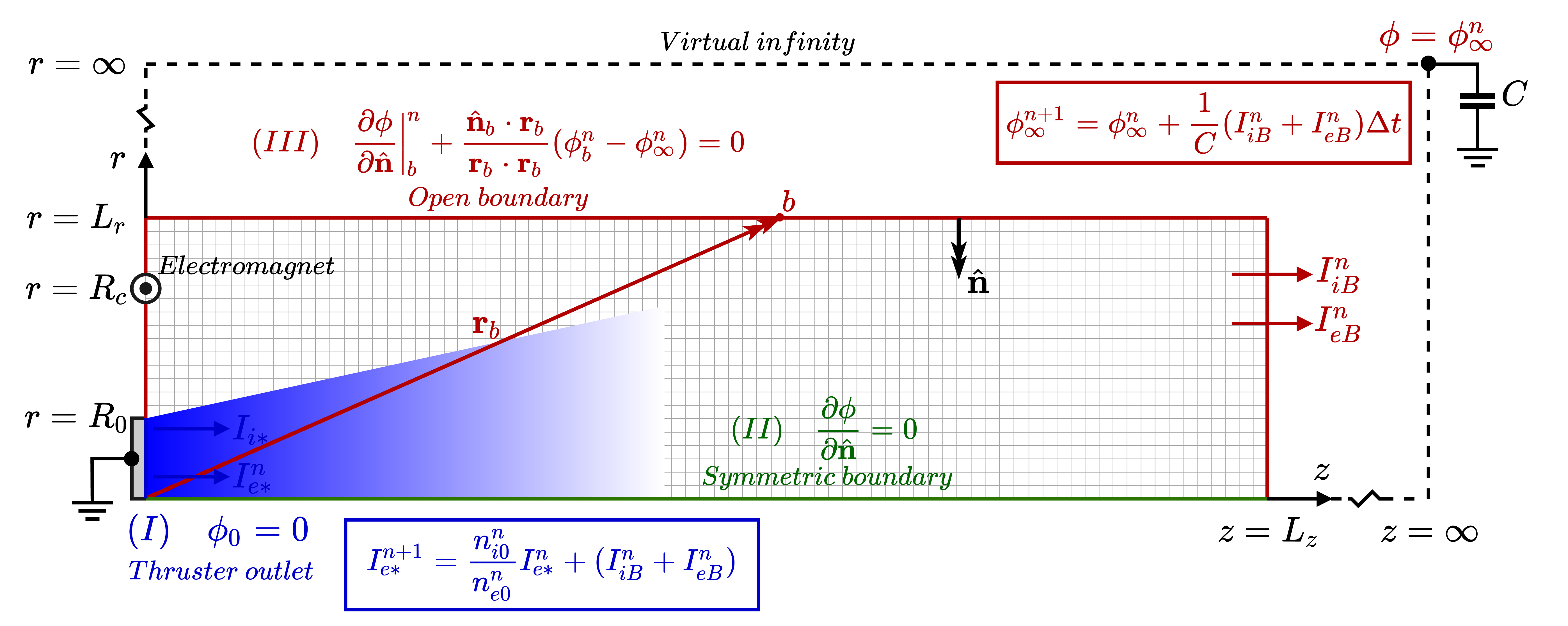}
    \caption{ ($I$) Thruster outlet, ($II$) symmetric boundary, and ($III$) open boundary. The electromagnet producing the MN is indicated at $r=R_c$. For the Poisson's equation, a Dirichlet condition applies to ($I$), zero-Neumann to ($II$), and Robin to ($III$). Ions and neutrals are absorbed on ($I$) and ($III$). Electrons are absorbed on ($I$) and selectively reflected on ($III$) based on an energy criterion. All particles are specularly reflected on ($II$). $I_{i*}$ and $I_{e*}$ are the injected ion and electron currents. $I_{iB}$ and $I_{eB}$ are the ion and electron currents lost to ($III$). $\phi_\infty$ is the free-space potential at infinity.}
    \label{fig:domainBC}
\end{figure*}

The model has been developed by adapting {the fully kinetic 2D-axisymmetric PIC code} \textit{Starfish}, which has been used previously to model Hall thruster channels \cite{b:brieda2012}, ion thruster plumes \cite{b:iac2019} and the plume of a magnetically-enhanced vacuum arc thruster \cite{b:zolotukhin2020}. An overview of the simulation domain is shown in \fref{fig:domainBC}, {consisting of a cylindrical 2D region $(z,r)$. The plasma source has not been included, since the scope of this work is to simulate purely the plume expansion.} Instead, ions, electrons and neutrals are injected through a boundary corresponding to the thruster outlet ($I$). The external boundaries ($III$) are treated as open to vacuum, connected to the thruster outlet ($I$) via a virtual free-space capacitance which ensures equal ion and electron current streams to the infinity at steady-state. Boundary ($II$) is the axis of symmetry.

{Hereafter, the subscripts $*$, $0$, $b$ and $\infty$ shall refer to properties within the plasma source (reference), at the thruster outlet boundary ($I$), at the open boundaries ($III$) and at the virtual infinity respectively. {The subscript $B$ shall refer to the integral sum of local properties along the open boundaries.} Likewise, the superscripts $+$ and $-$ shall refer to the forward and backward-marching components of the plasma properties.}

\subsection{Particle-in-Cell simulation}
\label{sec:pic}
{The set of macro-particles $p = 1,...,N_p$ with positions $\mathbf{r}^n_p=\langle z^n_p,r^n_p,0\rangle$, velocities $\mathbf{v}^n_p=\langle v_{pz}^n,v^n_{pr},v^n_{p\theta}\rangle$, masses ${m}_p$, and charges $q_p$, describe the ion $i$, electron $e$ and neutral $g$ dynamics at the $n$-th time-step $t^n = n\Delta t$. The particle motion is solved explicitly with the standard leap-frog Boris algorithm \cite{b:birdsall2005},}
\begin{eqnarray}
 & \frac{\mathbf{v}^{n+1/2}_p -\mathbf{v}^{n-1/2}_p}{\Delta t}=\nonumber\\  &\qquad \frac{q_p}{m_p}\left( \mathbf{E}^n(\mathbf{r}_p^n)+\frac{\mathbf{v}_p^{n+1/2}+\mathbf{v}_p^{n-1/2}}{2} \times \mathbf{B}(\mathbf{r}_p^n)\right) 
 \end{eqnarray}
\begin{equation}
    \frac{\mathbf{r}^{n+1}_p-\mathbf{r}^{n}_p}{\Delta t}=\mathbf{v}^{n+1/2}_p, \qquad p = 1,...,N_p,
\end{equation}
where $\mathbf{E}^n$ is the electric field, and $\mathbf{B}$ is the static background magnetic field. The movement of particles to new positions leads to a new distribution of charge density $\rho=n_i-n_e$, where $n_i$ is the ion density and $n_e$ is the electron density. It is computed by scattering particles to the mesh nodes using a second-order Ruyten shape factor $S_p$ \cite{b:ruyten1993},
\begin{equation}
    \rho = \frac{1}{{V}}\sum_{p=1}^{N_p}{q_p w_p S_p},
    \label{eqn:rho}
\end{equation}
where $w_p$ is the macro-particle specific weight \cite{b:birdsall2005}, and ${V}$ the mesh cell volume. The charge density is then used to solve for the self-consistent plasma potential $\phi$ according to the Poisson's equation, using an explicit successive over-relaxation (SOR) Gauss-Seidel scheme \cite{b:briedaThesis}.
\begin{equation}
    \epsilon_0\nabla^2\phi=-\rho,
\end{equation}
where $\epsilon_0$ is the vacuum permittivity. The electric field $\mathbf{E}=-\nabla\phi$ is then updated for the next time-step. To comply with typical PIC stability criteria, the mesh spacing is kept below the expected Debye length $\lambda_D = \sqrt{\epsilon_0k_BT_e/n_ee^2}$, where $k_B$ is the Boltzmann constant and $e$ is the elementary charge.

In order to reduce the computational burden, a numerical acceleration scheme has been adopted. The vacuum permittivity is increased by a factor $\gamma^2$ and the mass of heavy species reduced by a factor $f$. {This method does not require a scaled increase of the magneto-static field \cite{b:chen2020}, which is particularly useful when handling a MN expansion so as not to impose intractable conditions on the time-step required to resolve the electron gyroperiod $\omega_{ce}=e|\mathbf{B}|/m_e$.} The relationships between the simulated and physical constants, provided by $f$ and $\gamma$, are shown in equation~\eref{eqn:scale}. The direct consequences of these factors on $\lambda_D$ and plasma frequency $\omega_{pe}=\sqrt{n_ee^2/m_e\epsilon_0}$ are provided in equation~\eref{eqn:scale2}.
\begin{equation}
\tilde{m}_{i,n} \equiv \frac{{m_{i,n}}}{f} \qquad \tilde{\epsilon}_0 \equiv \gamma^2\epsilon_0
\label{eqn:scale}
\end{equation}
\begin{equation}
\tilde{\lambda}_D = \gamma\lambda_D \qquad \tilde{\omega}_{pe} = \frac{\omega_{pe}}{\gamma}
\label{eqn:scale2}
\end{equation}
{where scaled quantities have been referred to with the diacritic $\sim$.} A summary on the scaling relationship of other parameters and retrieving physical results can be found in reference \cite{b:szabo}. 

Collision processes between ions, electrons and neutrals are simulated using a combination of Direct Simulation Monte Carlo (DSMC) \cite{b:bird1998} and Monte Carlo Collision (MCC) \cite{b:birdsall1991} methods. All simulations contained within this article consider seven different collision processes: electron-electron Coulomb scattering, electron-ion Coulomb scattering, electron-neutral elastic scattering, electron-neutral ionisation, ion-neutral elastic scattering, ion-neutral charge exchange, and neutral-neutral elastic scattering. All relevant cross-sections are recovered from the LXCat database \cite{LXCat}. 
\subsection{Boundary conditions on the Poisson's equation}
\label{sec:robin}

The thruster outlet ($I$) is given the reference potential $\phi_0=0$. The $r=0$ boundary ($II$) is symmetric, therefore the zero-Neumann condition $\partial\phi / \partial \hat{\mathbf{n}}=0$ is applied there. {Concerning the open boundaries ($III$), a non-stationary Robin-type condition is introduced of the form
\begin{equation}
    \frac{\partial\phi}{\partial\hat{\mathbf{n}}}\bigg|^n_b+\frac{\hat{\mathbf{n}}_b\cdot{\mathbf{r}_b}}{\mathbf{r}_b\cdot\mathbf{r}_b}\left(\phi^n_b-\phi^n_\infty\right) = 0
    \label{eqn:robin}
\end{equation}
where $\mathbf{r}_b=\langle z_b-z_0, r_b-r_0, 0\rangle$ is the vector distance from the centre of the thruster outlet ($I$) to the location on the open boundary ($III$), $\mathbf{\hat{n}}_b$ is the inward-pointing unit normal, and $\phi_\infty$ denotes the free-space plasma potential at infinity.}

\Eref{eqn:robin} is a transparent condition modelling the $1/r$ monopole decay of the potential \cite{b:griffiths2013} far into the plasma $(\phi\rightarrow\phi_\infty, {\partial\phi}/{\partial\hat{\mathbf{n}}}\rightarrow0, \phi_\infty<0)$; its derivation is provided in the Appendix. {This condition inherently depends on the value of the potential drop across the plume $|\phi_\infty|$.} {The procedure adopted to self-consistently calculate $\phi_\infty$ is described in \sref{sec:cap}. Finally, it is worth noting that the condition expressed in equation~\eref{eqn:robin} is a generalisation of the commonly used zero-Neumann assumption \cite{b:chen2020} for the case of a finite domain.}

\subsection{Boundary conditions on particle kinetics}
\label{sec:electrorefl}

{At each time-step, ions, electrons and neutrals are injected from the thruster outlet boundary ($I$). For all species, a Maxwellian VDF is assumed \cite{chen1999upper}, with reference temperatures $T_{k*}$ for $k=i,e,g$. {A drift velocity equal to the Bohm speed is imposed along the $z$ direction for both ions and electrons $\mathbf{u}_{i,e} = \langle c_{*},0,0\rangle$, where $c_{*}=\sqrt{k_B T_{e*}/m_i}$ \cite{b:martinez2015}. Neutrals possess diffusion drift velocity $\mathbf{u}_{g} = \langle \bar{v}_g/4,0,0\rangle$, where $\bar{v}_g=\sqrt{8k_BT_{g*}/\pi m_g}$ \cite{b:magarotto2020}}. The resultant generic VDF reads
\begin{eqnarray}
& f_{k}^{+}(\mathbf{v}_{k}) = \sqrt{\frac{m_{k}}{2\pi T_{k*}}}\exp \left(-\frac{m_{k}}{2T_{k*}}|\mathbf{v}_{k}-\mathbf{u}_{k}|^2 \right)H(v_{kz}) \nonumber\\ &~
\label{eqn:f}
\end{eqnarray}
The Heaviside $H$ makes the distribution one-sided, since only forward-marching distributions ($v_{kz}>0$) can be imposed. The backward-marching distributions are an output of the simulation, which strongly depends on the steady-state value of $\phi_\infty$ (see \sref{sec:cap}). In this regard, any backward-marching electrons (and ions or neutrals) returning to the thruster outlet ($I$) are absorbed. In fact, once inside the source where the density is high and collisions are frequent, a particle will become re-equilibrated with the source plasma and lose all its memory when it is re-injected into the beam \cite{b:lafleur2015,chen1999upper}. The values of $T_{k*}$ represent the expected plasma properties within the source; they are not necessarily equal to the final temperature at the thruster outlet $T_{k0}$ since the dynamics of the backward-marching species are not known \textit{a priori} (e.g., $f_{e}^{-}$ might be non-Maxwellian).} At the symmetry plane ($II$), all particles are specularly reflected.

Since ions are accelerated outward by the ambipolar electric field \cite{b:lafleur2015}, no special treatment is required at the open boundaries ($III$); therefore ions reaching them are simply absorbed. Neutrals are also absorbed. For electrons, the behaviour is not as straightforward. Generating a stable steady-state plume without altering the electron kinetics requires an energy-based treatment \cite{b:li2019}. {Physically, two separate populations of electrons can be identified depending on their total energy, namely trapped and free. The former are the less energetic electrons that cannot overcome the potential drop that occurs across the plume. The trapped electrons are forced to turn back to the plasma source at a certain distance downstream. The free population are the electrons that do have energy enough to cross the potential drop and thus escape to infinity. Assuming a steady-state, and axisymmetric electric field, the total energy of each electron can be defined as 
\begin{equation}
    E_e=\frac{1}{2}m_e(v_{ez}^2+v_{er}^2+v_{e\theta}^2)-e\phi(\mathbf{r})
    \label{eqn:energy}
\end{equation}
where $E_e$ is a constant conserved quantity of the motion in the collisionless case. From energy conservation, trapped electrons are characterised by $E_e<|e\phi_\infty|$, while for free electrons $E_e\geq|e\phi_\infty|$.}
{From these considerations, the following boundary condition is defined.} When an electron reaches an open boundary node $b$, it's kinetic energy is taken as $KE_{eb}=\frac{1}{2}m_e|\mathbf{v}_{eb}|^2$, and then compared to the trapping potential {$PE_{b}=e(\phi_b-\phi_\infty)$}. 
\begin{itemize}
    \item If $KE_{eb}<PE_{b}$ the electron is trapped, so it is reflected back with velocity $-\mathbf{v}_{eb}$.
    \item Else, it is a free electron to be removed from the domain.
\end{itemize}  
This boundary condition therefore allows the highest-energy electrons to escape, but retains the physical proportion of the trapped population, ensuring stability \cite{b:li2019,b:brieda2018}. Finally, the key assumptions in this energy-based boundary condition are, firstly, that the plasma is collisionless downstream of the open boundaries and that, beyond the domain, the magnetisation is not so strong as to induce reflection of highly energetic electrons \cite{b:martinez2015}.

\subsection{Capacitive circuit} 
\label{sec:cap}

{The value of $\phi_\infty$ is a non-stationary unknown and must be calculated self-consistently as part of the solution. From the energy-based criterion discussed in \sref{sec:electrorefl}, there is a value of $\phi_\infty$ that reflects sufficient electrons to maintain a current-free plume. Therefore, the value of $\phi_\infty$ is self-consistently controlled via a virtual free-space capacitance $C$. The resultant control algorithm reads}
\begin{equation}
    \phi_\infty^{n+1}=\phi_\infty^n+\frac{1}{C}(I_{iB}^n{f}^{-0.5}+I_{eB}^n)\Delta t,
    \label{eqn:control}
\end{equation}
where $I_{iB}$ and $I_{eB}$ are the sum ion and electron currents leaving the open boundaries ($III$), with the factor ${f}^{-0.5}$ scaling down the ion current in accordance with the applied mass factor \cite{b:szabo}. The value of $C$ must be carefully chosen according to a compromise between fast convergence of $\phi_\infty$ and stability of the Poisson's solver {(see the sensitivity analysis reported in \sref{sec:steady}).} This method ensures that the system evolves self-consistently and inherently guarantees that, once at steady-state, the ion and electron currents are equal ($I_{iB}=-I_{eB}$) at the open boundaries ($III$), and therefore also at the infinity.
The initial value of $\phi_\infty$ {($\phi_\infty^0$)} is set according to the theoretical value obtained by assuming a current-free condition at the thruster outlet, the absence of a magnetic field, and electron energy conservation \cite{b:difede2021}. For the Maxwellian population given by equation~\eref{eqn:f}, the analytical result is
\begin{eqnarray}
 &\frac{4u_{i0}}{\bar{v}_{e*}}=1+erf  \sqrt{\frac{-e\phi_\infty^0}{T_{e*}}} -\sqrt{\frac{-2e\phi_\infty^0}{\pi T_{e*}}}exp\left(\frac{e\phi_\infty^0}{T_{e*}}\right) \nonumber\\ &~
\label{eqn:phi0}
\end{eqnarray}
where $\bar{v}_{e*}=\sqrt{8k_BT_{e*}/\pi m_e}$ is the mean reference electron velocity. For $u_{i0}=c_{*}$ with xenon, equation~\eref{eqn:phi0} yields $e\phi_\infty^0\sim -6.4T_{e*}$.

{Controlling only the value of $\phi_\infty$ by means of equation~\eref{eqn:control} is not in itself sufficient to implement a self-consistent circuit condition. Any non-zero net current in the transient leaving the open boundaries ($III$) must be re-injected into the domain via the thruster outlet ($I$) \cite{b:brieda2018}. Moreover, the injected electron current $I_{e*}$ is controlled in order to enforce the quasi-neutrality condition at the thruster outlet ($I$), namely
\begin{equation}
n_*=n_{i0} = n_{e0}^+ + n_{e0}^-
\label{eqn:quas-neutr}
\end{equation}
Being $n_{e0}^-$ unknown \textit{a priori}, equation~\eref{eqn:quas-neutr} can be satisfied at the steady-state by adjusting $I_{e*}$ and, in turn, $n_{e0}^+$. From these considerations, the following conditions are imposed to the particles injected from boundary ($I$). Ions are injected with a constant current given by $I_{i*} = en_{*}c_{*}A_0$, where $A_0$ is the area of the thruster outlet. The injected electron current is updated each time step according to 
\begin{equation}
{I_{e*}^{n+1}} = (I_{iB}^n+I_{eB}^n) + \frac{n_{i0}^n}{n_{e0}^n}{I_{e*}^n}
\label{eqn:QN}
\end{equation}
where the first term completes the circuit and the second enforces the quasi-neutrality. This condition guarantees that quasi-neutrality and current-free conditions are respected at the steady-state. A similar control strategy has not been imposed to ions since $I_{i* }\equiv I_{i0} \equiv I^+_{i0}$, whereas $I_{e*} \equiv I_{e0}^+ \neq I_{e0}$. Considering that injected electrons are Maxwellian, the initial value of the current is set as ${I_{e*}^0}=-en_{*}(\bar{v}_{e*}/4+c_{*})A_0$.} The neutral flux is imposed as $\Gamma_{g*}=n_{g*}\bar{v}_g/4A_0$.

It is prudent to state that $\phi_\infty$ and $I_{e*}$ are numerically stored as their moving average, therefore minimising any PIC noise from the fluctuations in their value. There thus exists a fully-consistent relationship between the current flowing from the plasma source $I_{i0}=-I_{e0}$ to the open boundaries $I_{iB}=-I_{eB}$, the potential drop $\phi_\infty$ and the macroscopic plume solution. It is established via the boundary condition of equation~\eref{eqn:robin}, the electron energy reflection condition, and the capacitive circuit control of equation~\eref{eqn:control} and equation~\eref{eqn:QN}.

\section{Verification of the numerical model}
\label{sec:verif}

\begin{table*}[!htb]
\caption{\label{tab:verparams}  Simulation parameters.}
\lineup
\centering
\begin{indented}
\item[]\begin{tabular}{@{}llll@{}}
\br
Parameter                          &                               & Unscaled             & Scaled$^b$              \\ \mr
Thruster Outlet Radius               & $R_0$ [mm]                  & 7                   & -                    \\
Ion Mass (Xe)                      & $m_i$ [kg]                  & $2.18\times10^{-25}$ & $5.45\times10^{-23}$ \\
Propellent Mass Flow Rate          & $\dot{m}$ [mg/s]        & 0.15                & 2.37                \\
Reference Plasma Density           & $n_{*}$ [m$^{-3}$]              & $1.6\times10^{18}$     & -                    \\
Reference Neutral Density & $n_{g*}~[m^{-3}]$& $5.3\times 10^{19}$ & - \\
Reference Ion Temperature          & $T_{i*}$ [K]               & 298                  & -                    \\
Reference Electron Temperature     & $T_{e*}$ [eV]              & 5                    & -                    \\
Reference Neutral Temperature     & $T_{g*}$ [K] & 298 & - \\
Reference Bohm Speed              & $c_{*}$ [m/s]         & 1920                 & 30310                \\
Ion Current                        & $I_{i*}$ [A]                & 0.0755             & 1.19               \\
Electromagnet Radius                & $R_c$ [mm]                  & 25.2                   & -                    \\
Throat Magnetic Field Strength     & $B_0$ [G]                   & $0-1200$              & -                    \\ \mr
Axial Domain Length                & $L_z$ [m]                  & 0.250                  & -                    \\
Radial Domain Length               & $L_r$ [m]                  & 0.100                  & -                    \\
Simulated Time                     & $\tau$ $[\mu s]$                 & -                    & 80                   \\
Reference Debye Length             & $\lambda_{D*}$ [mm]        & 0.0131             & $0.350$              \\
Reference Plasma Frequency         & $\omega_{pe*}$ [rad/s] & $7.17\times10^{10}$  & $2.68\times10^{9}$   \\
Electron Gyro-Frequency$^a$          & $\omega_{ce0}$ [rad/s]  & $1.76\times10^{10}$  & -                  \\
Number of Axial Cells              & $N_z$                         & -                    & 500                  \\
Number of Radial Cells             & $N_r$                         & -                    & 200                  \\ 
Number of Time-steps$^a$           & $N_t$                         & -                    & $6.4\times10^6$     \\
Steady-state Macro-particles               & $N_p$                         & -                    & $\sim 8\times10^5$    \\
\br
\end{tabular}
\item[] $^a$ $100G$.
\item[] $^b$ $f=250$, $\gamma=26.7$.
\end{indented}
\end{table*}

{This section demonstrates the robustness of the new boundary conditions. The verification is divided into two parts. First the steady-state stability is demonstrated against the classical open-outflow boundary conditions \cite{b:levin2018, b:hu2017}. Namely, constant electron current injected at the plasma source, absorption of all electrons reaching the open boundaries ($III$), and the zero-Neumann condition on the Poisson's equation.} {Second, a domain independence study is presented, evaluating both macroscopic plasma parameters and the propulsive performance (i.e. thrust) with respect to both axial and radial domain dimensions.}

\begin{figure}[!htb]
    \centering
    \includegraphics[width=0.98\linewidth]{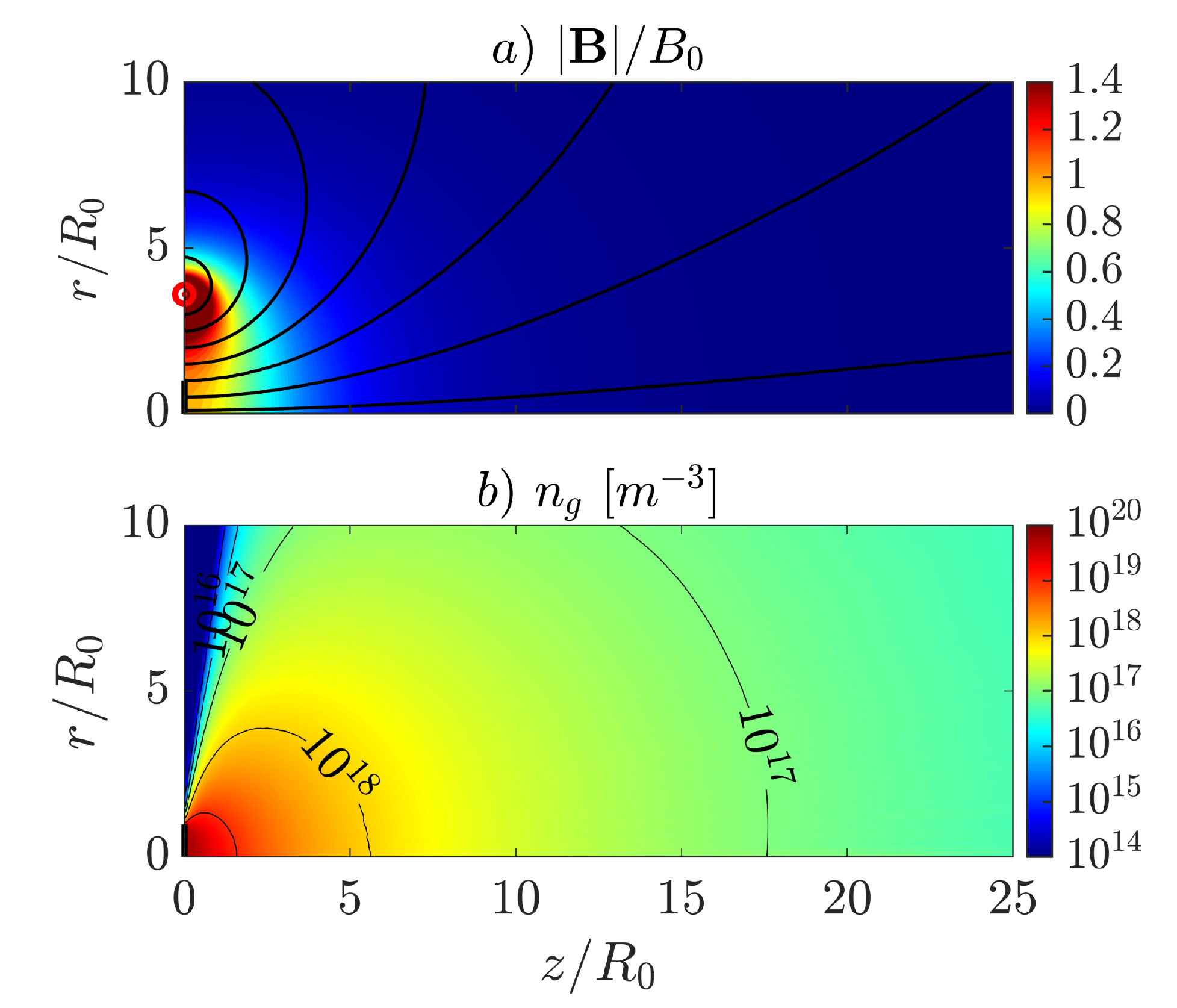}
    \caption{(a) Normalised magnetic field magnitude $|\mathbf{B}|/B_0$. The electromagnet has radius $R_c=3.6R_0$. (b) Neutral density $n_g$ from the DSMC simulation.}
    \label{fig:b_field}
\end{figure}

\Tref{tab:verparams} summarises the physical and numerical parameters which are used, unless otherwise specified, throughout this article. {Xenon is the propellant gas, with the reference plasma properties assumed within the source typical of the operating conditions in a low-power (50~W) HPT \cite{b:manente2019,b:magarotto2020}. A purely divergent MN is produced by an electromagnet of radius $R_c$, positioned concentric with the thruster outlet of radius of $R_0$ {\cite{pottinger2011performance}}.} {\Fref{fig:b_field}(a) illustrates the magnetic field topology $\mathbf{B}$ on the nominal simulation domain, normalised with its value at the magnetic throat, that is $B_0=|\mathbf{B}(0,0)|$. Before commencing the PIC simulations, the DSMC method was used to pre-compute the neutral gas density field $n_g$ given in \fref{fig:b_field}(b).} 

{The scaling factors applied to the ion/neutral mass and the vacuum permittivity are $f=250$ and $\gamma=26.7$ respectively. The latter value is chosen such that the thruster outlet ($I$) is resolved with 20 cells $R_0=20\tilde{\lambda}_{D*}$. The domain spans $L_z=25R_0$ in length and $L_r=10R_0$ in height, with a uniform mesh spacing of $\tilde{\lambda}_{D*}$. The time-step adopted satisfies $\Delta t \omega_{ce0}=0.35$. In this way, the electron gyro-motion is resolved in all the domain and the stability criterion on the resolution of the scaled plasma frequency \cite{b:birdsall2005} is also satisfied (see \tref{tab:verparams}). The Poisson's equation is solved every $0.286\tilde{\omega}_{pe*}/\omega_{ce0}$ iterations \cite{b:birdsall2005}. Neutrals are sub-cycled at a larger time-step according to their Courant–Friedrichs–Lewy (CFL) condition \cite{b:bird1998}.} Macro-particle weights are selected so as to maintain an average number per cell above 10 at the steady-state, {as a result $N_p\approx8\times10^5$.} The steady-state is characterised by the number of macro-particles leaving the domain matching the number of newly injected macro-particles at the thruster outlet ($I$) within 0.01\% for a defined number of iterations. On a machine equipped with an Intel\textsuperscript{\textregistered} i7-7700 @3.6~GHz $\times$ 8, and 32~Gb of RAM, the computational time is approximately 9.6~hrs to reach steady-state. Approximately a further 8~hrs is required for steady-state averaging over 50000 time-steps.   
\subsection{Steady-state stability}
\label{sec:steady}

\begin{figure}[!htb]
    \centering
    \includegraphics[width=\linewidth]{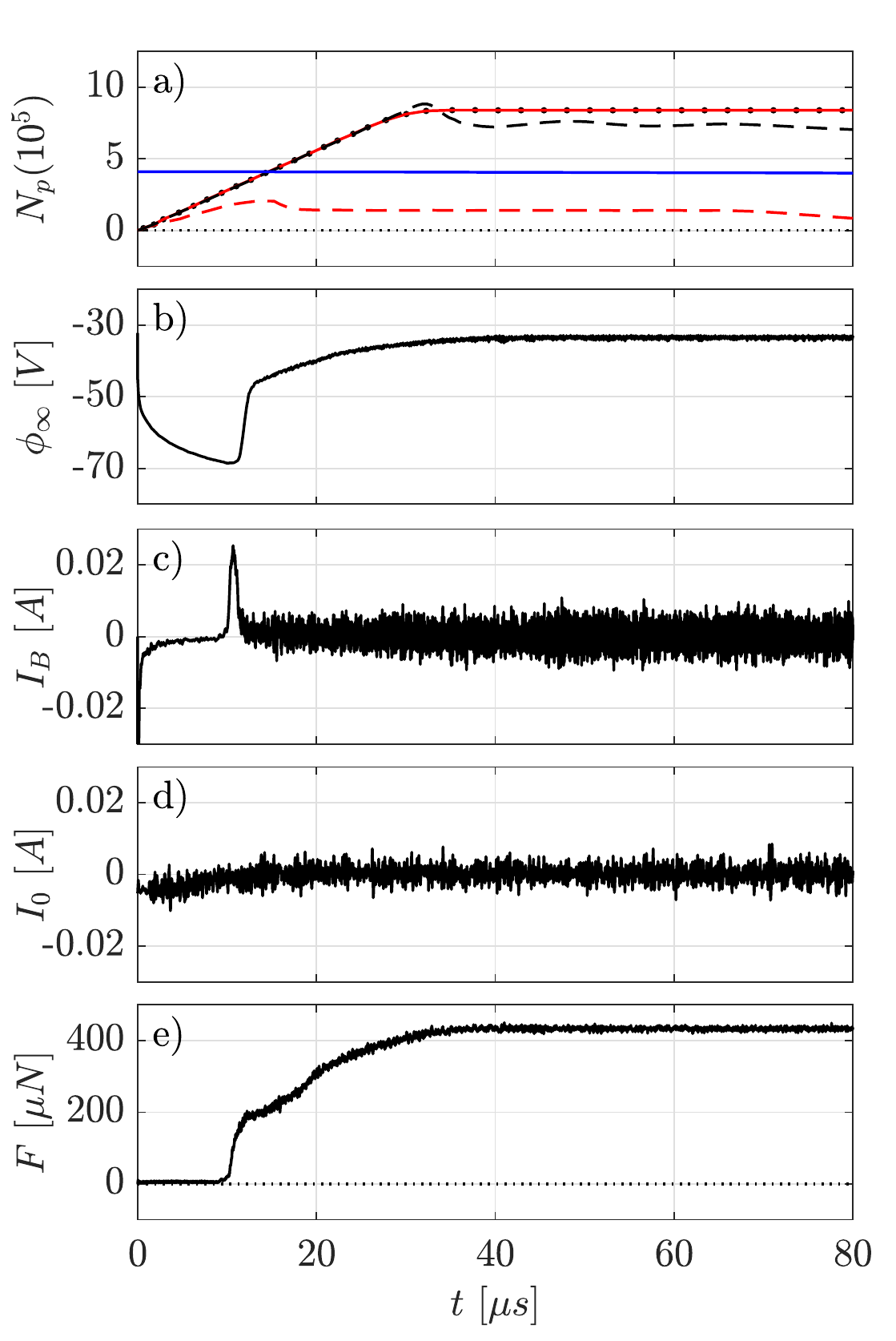}
    \caption{Transient evolution of (a) Total number of macro-particles $N_p$ for ions (\fullcircle), electrons (\textcolor{red}{\full}) and neutrals (\textcolor{blue}{\full}) in the new model, and ions (\dashed) and electrons (\textcolor{red}{\dashed}) for open outflow conditions; (b) free-space potential $\phi_\infty$; (c) net current $I_B$ at the open boundaries ($III$); (d) net current $I_0$ at the thruster outlet ($I$); and (e) thrust $F$.}
    \label{fig:oneDtest}
\end{figure}
{The stability of the new model is assessed for the unmagnetised case (i.e. $B_0=0$), since results can be more easily compared to theoretical values}. The time-step is defined by $\Delta t \omega_{pe*}=0.05$. \Fref{fig:oneDtest}(a) gives the evolution of the macro-particle count for both the new model and the open-outflow conditions. Although both simulations start with a similar growth in ion count during the transient, the electron count peaks around $14$~$\mu$s with the open-outflow boundary. Despite the number of ions continuing to increase, the electron count decreases, resulting in a growing charge imbalance. This eventually results in the formation of a virtual anode \cite{b:brieda2018,b:briedaThesis} around $32$~$\mu$s, which is followed by gradual ion loss. The vast majority of electrons are lost, thus the simulation collapses; this is the ``numerical pump instability'' \cite{b:brieda2018,b:briedaThesis}. No such instability is observed with the new model. The electron and ion populations closely trend each other. Steady-state is achieved near $34$~$\mu$s, and the ion and electron counts remain invariant for the reminder of the simulation. There is negligible change in the neutral count.

To prove the new model can self-consistently calculate the free-space potential $\phi_\infty$, \fref{fig:oneDtest}(b) plots its value against the simulation time. The trend mimics the voltage seen across the charging-discharging cycle of a capacitor \cite{b:griffiths2013}. Indeed, from \fref{fig:oneDtest}{(c)}, an initially large negative net current $I_B=I_{iB}+I_{eB}$ at the open boundaries ($III$) is clear. Therefore $|\phi_\infty|$ increases (the charging cycle) to slow down fast electrons. The rate of increase slows (see \fref{fig:oneDtest}(b)) as fewer electrons can escape the growing potential barrier, and $I_B$ becomes negligible towards $10$~$\mu$s (see \fref{fig:oneDtest}{(c)}). The minimum of the voltage curve, at around $10$~$\mu$s, represents the time at which ions begin to cross the open boundaries ($III$). $|\phi_\infty|$ begins to decrease (the discharging cycle) as fewer electron reflections are required to balance a now net positive current that peaks at $11$~$\mu$s (see \fref{fig:oneDtest}{(c)}). After this initial recovery, a further, but slower, decrease in $|\phi_\infty|$ occurs as the ion beam current ---which is itself determined by the ambipolar acceleration of $|\phi_\infty|$--- establishes an equilibrium state. After $39$~$\mu$s, $I_B$ fluctuates about zero and a steady-state value of $\phi_\infty=-33.5$~V is reached. This is similar to the theoretical initial value of $\phi_\infty^0=-32.2$~V; it also falls between the values of $-28.9$~V and $-37.5$~V given by alternatives to equation~\eref{eqn:phi0} in references \cite{b:lafleur2011} and \cite{b:merino2018kinetic} respectively.

At the steady-state, a zero net current $I_0=I_{i0}+I_{e0}$ is also achieved across the thruster outlet boundary ($I$) as shown in \fref{fig:oneDtest}{(d)}. This arises purely as a product of the self-consistent electric field coupling electrons to the ion beam, since the current-free condition is only enforced at the open boundaries ($III$); only quasi-neutrality is enforced at ($I$). After an initially net negative current, caused by the re-injection of electrons as per the circuit condition, steady-state is achieved at around $8$~$\mu$s. Lastly, \fref{fig:oneDtest}(e) provides the convergence of thrust, which achieves steady-state in the same time as $\phi_\infty$ at $39$~$\mu$s. The steady-state thrust is $F=431$~$\mu$N.

\begin{figure}[!tb]
    \centering
    \includegraphics[width=\linewidth]{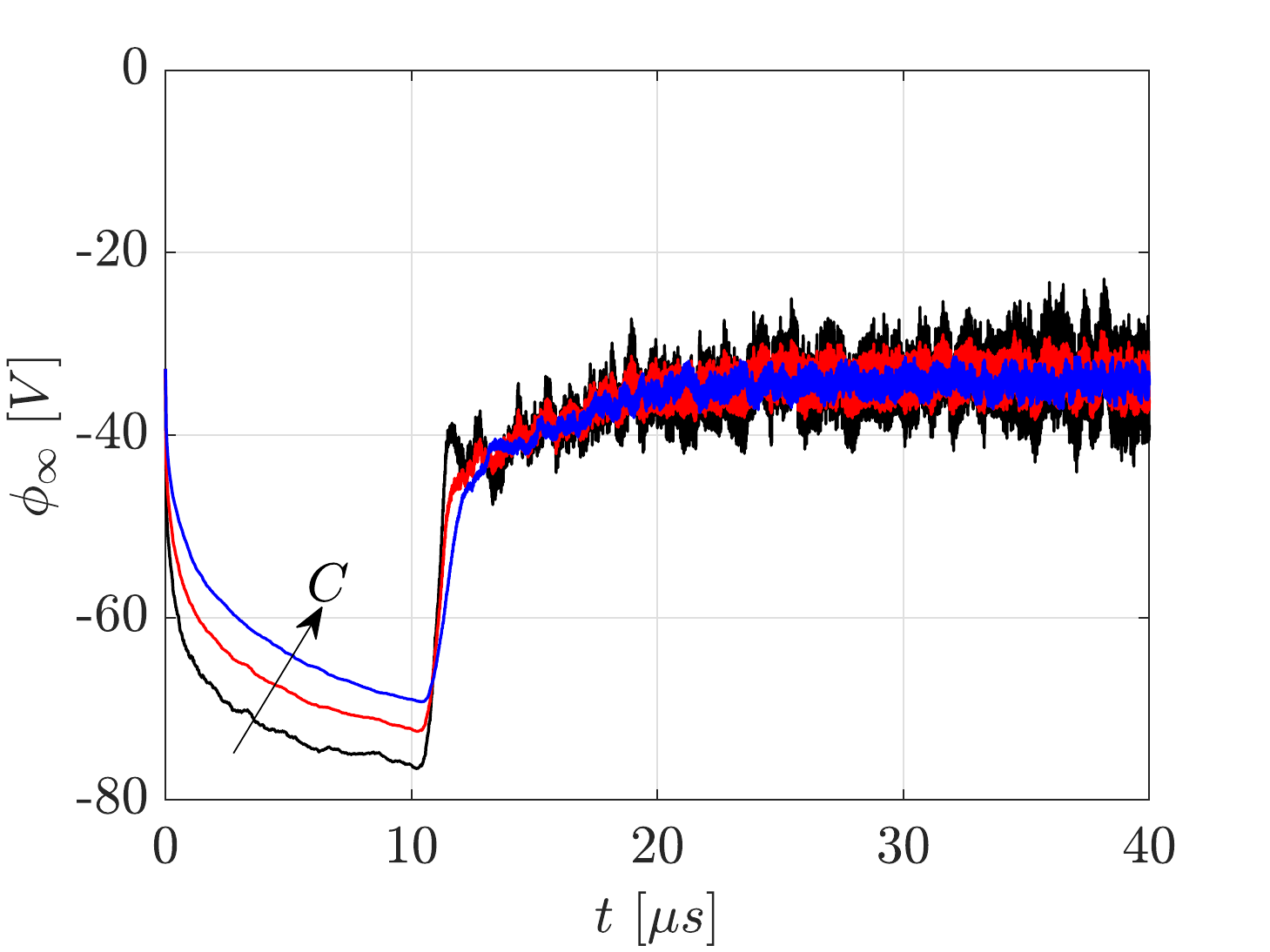}
    \caption{Evolution of the plasma potential at infinity $\phi_\infty$ for three values of the virtual free-space capacitance $C$: $0.2$~nF (\full); $0.4$~nF (\textcolor{red}\full); and $0.8$~nF (\textcolor{blue}\full).}
    \label{fig:C_independance}
\end{figure}
\begin{figure*}[!htb]
    \centering
    \includegraphics[width=\linewidth]{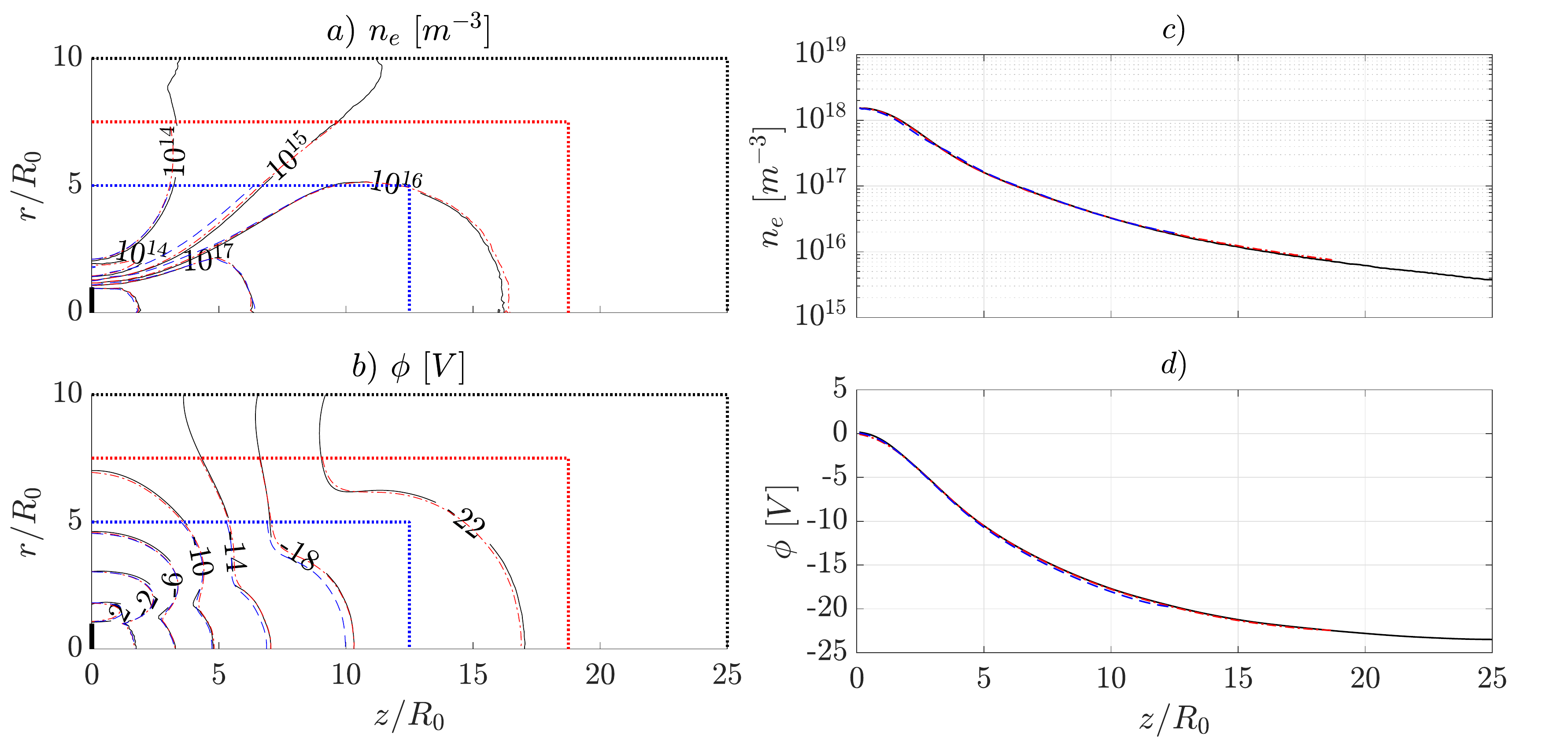}
    \caption{(a) Electron density $n_e$ and (b) plasma potential $\phi$ using the nominal $25R_0 \times 10R_0$ (\tikz[baseline]{\draw[thick] (0,.5ex)--++(.5,0) ;}), three-quarter $18.75R_0 \times 7.5R_0$ (\textcolor{red}{\tikz[baseline]{\draw[dash dot] (0,.5ex)--++(.5,0) ;}}) and half $12.5R_0 \times 5R_0$ ({\textcolor{blue}{\tikz[baseline]{\draw[dashed] (0,.5ex)--++(.5,0) ;}}}) domains. Dotted lines (\tikz[baseline]{\draw[dotted] (0,.5ex)--++(.5,0) ;}) indicate the boundaries of the reduced domains. {(c) Electron density $n_e$ and (d) plasma potential $\phi$ sampled on the symmetry axis $r=0$.}}
    \label{fig:domain_ind}
\end{figure*}

{The results obtained with different values of the virtual free-space capacitance $C$ have been compared in \fref{fig:C_independance} in terms of $\phi_\infty$. Firstly, all three cases converge to a similar steady-state value within an accuracy of $0.4$~V. Second, {for smaller values of $C$}, the voltage drop in the charging cycle increases: around $-69.0$, $-72.4$ and $-76.4$~V for {$C=0.8$, $0.4$ and $0.2$~nF respectively.} This marginally increases the ambipolar acceleration of the ions during the transient. Thus, the point at which ions begin to cross the open boundaries ($III$) occurs $0.26$~$\mu$s and $0.22$~$\mu$s earlier for {$C=0.2$ and $C=0.4$ compared to $C=0.8$~nF. Analogous to a simple capacitive circuit, an increase in the value of $C$ increases the rate of voltage drop, while also decreasing the discharging time and the voltage recovery. This results in the equilibrium state being achieved at approximately the same simulation time ($39$~$\mu$s) for all three cases}.} To conclude, while the value of $C$ affects the plume during the transient phase, the solution at steady-state is independent of it. This further confirms the robustness of the proposed simulation strategy.

{It is also important to note that smaller values of $C$ cause an increase in noise.} This is expected for a proportional-type control law (see equation~\eref{eqn:control}), since the value of $C$ places an effective resolution on the adjustment of $\phi_\infty$. It is therefore important to exercise care in the choice of virtual free-space capacitance. {$C$ must be small enough} to create the voltage drop necessary to prevent the instability, but not so small as to introduce additional noise to the solution and strain on the Poisson's solver. For the remainder of the simulations in this work, {$C=0.8$~nF is chosen.}

\begin{figure*}[!htb]
    \centering
    \includegraphics[width=\linewidth]{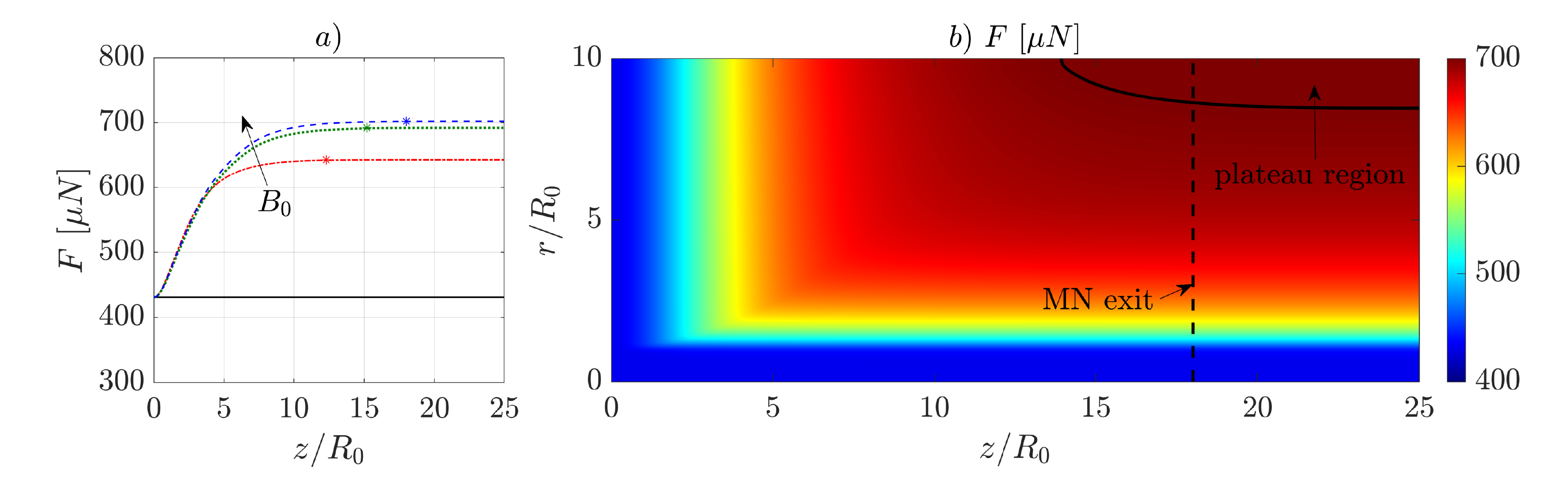}
    \caption{(a) Thrust $F$ for axial truncation of the domain for unmagnetised (\full), $100$~G (\textcolor{red}{\chain}), $300$~G (\textcolor{green}{\dotted}) and $600$~G (\textcolor{blue}{\dashed}) cases. The markers indicate the MN exit. (b) Thrust $F$ for axial and radial truncation in the $600$~G case. The solid contour (\full) represents the value of final thrust; the dashed line (\dashed) represents the MN exit.}
    \label{fig:axial_thrust}
\end{figure*}

\subsection{Domain independence}
\label{sec:domain}

For the case of $B_0=600$~G, simulations are performed to demonstrate the domain independence using domain sizes of $18.75R_0\times7.5R_0$ (three-quarter) and $12.5R_0\times5R_0$ (half) compared to the nominal $25R_0\times10R_0$. Both reduced domains converged to values of $\phi_\infty$ within $0.8$~V of the nominal $-39.1$~V. A comparison of the steady-state electron density and plasma potential distributions across the domains is shown in \fref{fig:domain_ind}(a) and \fref{fig:domain_ind}(b) respectively. It can be seen that the shape of the plume obtained from both reduced domain simulations are in very good agreement with the nominal case.

{For a more quantitative analysis, the electron density and plasma potential are compared along the axis of symmetry ($r=0$) in \fref{fig:domain_ind}(c) and \fref{fig:domain_ind}(d). Along the axis, results agree within 6\% for the electron density, and 2\% for the plasma potential. {The largest disagreement occurs outside the periphery of the outermost magnetic field line connected to the source, approximated by the $10^{15}$~m$^{-3}$ contour in \fref{fig:domain_ind}(a). The density in the reduced domain simulations is up to 68\% higher in this region compared to the nominal.} This difference may be attributed to the noise introduced due to the number of particles escaping and reflecting from the radial open boundary $(III)$, which is no longer significantly removed from the thruster outlet $(I)$. Nonetheless, the electron density within the core of the plume is in excellent agreement within 2\% on average. Overall, the comparisons demonstrate that the new set of boundary conditions provides a domain-independent solution within a tolerance smaller than the typical PIC noise.}

\begin{figure}[!tb]
    \centering
    \includegraphics[width=\linewidth]{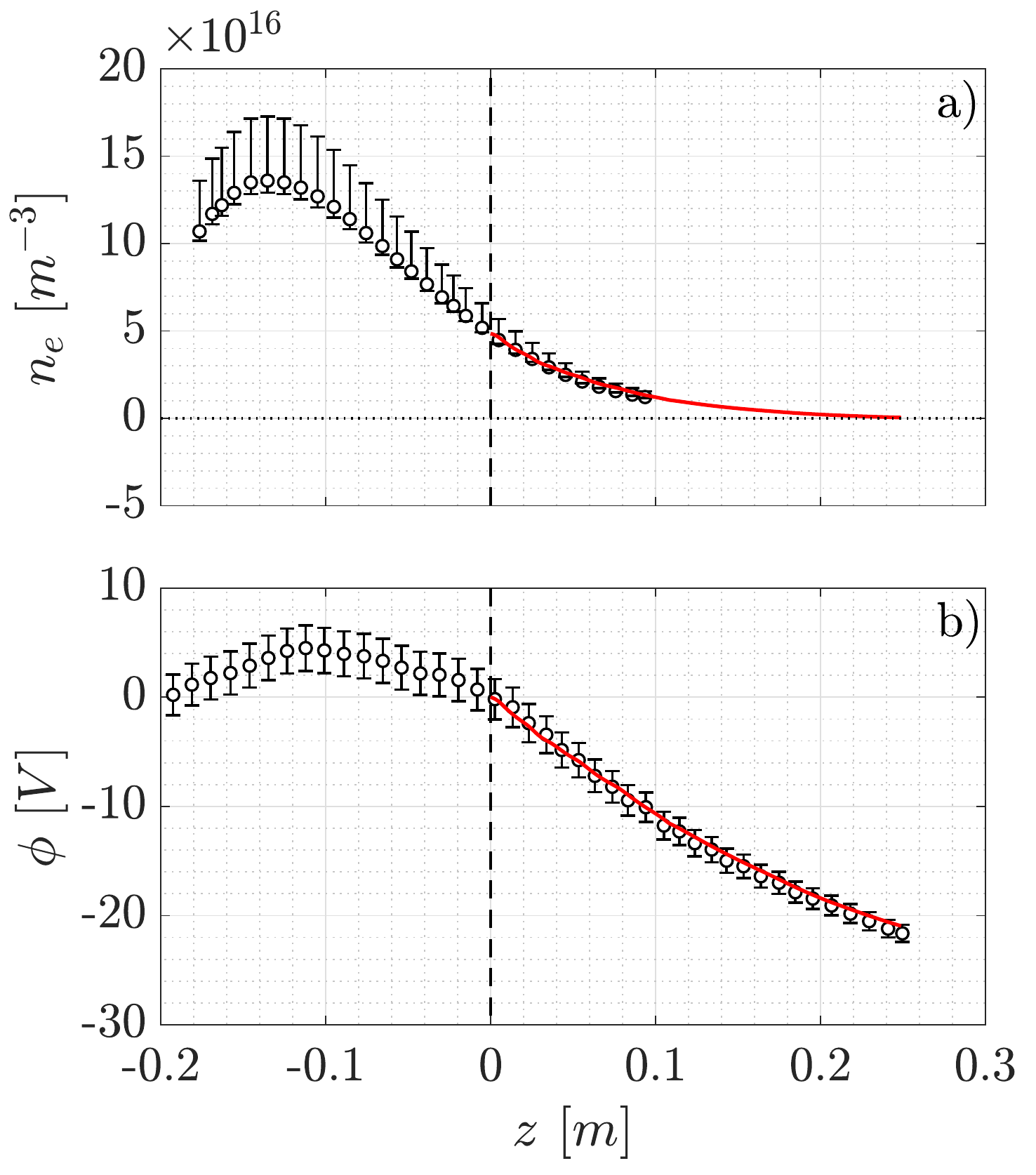}
    \caption{\centering Experimentally measured data (\opencircle) with relative uncertainty bands and PIC output (\textcolor{red}{\full}) on the axis of the \textit{Piglet} reactor: (a) electron number density $n_e$, and (b) plasma potential $\phi$.}
    \label{fig:valid_plot}
\end{figure}

The final choice of domain size depends primarily on the phenomenon of interest. If only plume-spacecraft interactions are desired, significant computational savings can be realised by applying the new boundary conditions to a comparatively small domain. However, if the thrust is required, a domain of sufficient axial length is needed. The thrust $F$ is given by the axial flux integral
\begin{equation}
F = \iint_{S_B} \sum_k \left(n_k m_k v_{kz}\mathbf{v}_k + p_k\hat{\mathbf{z}} \right) \cdot\hat{\mathbf{n}}~dS_B,
\label{eqn:thrust}
\end{equation}
where the two terms correspond to the species momentum flux and pressure $p_k=n_k k_B T_k$ respectively, for $k=i,e,g$. $S_B$ is the open boundary ($III$) surface.
{\Fref{fig:axial_thrust}(a) shows the thrust $F(z)$ calculated for different axial truncation of the domain. When the magnetic field is absent, the total axial flow momentum is conserved since no mechanism can exert force on the plasma; as a result $F(L_z)=F(0)=431$~$\mu$N.} In the $100$~G case, $F$ increases to a converged value of $643$~$\mu$N at $z\sim 12.3R_0$; for $300$~G, $F=692$~$\mu$N at $z\sim15.2R_0$; at $600$~G, $F=702$~$\mu$N at $z\sim18R_0$. This axial plane where the thrust establishes a plateau may be referred to as the exit of the MN, where plasma detachment occurs. The axial domain length must therefore include this plane so as to yield the accurate value of thrust. The domain size required is therefore proportional to the magnetic field strength.

The radial domain width must also be large enough. \Fref{fig:axial_thrust}(b) spatially maps the thrust $F(z,r)$ for both axial and radial truncation of the domain in the $600$~G case. {The size of the domain required to obtain a plateau value of $F$ is given by the area bounded by the solid contour. Accordingly, the domain must have an axial length $L_z\gtrsim 18R_0$ and radial width $L_r\gtrsim 8R_0$ to not underestimate the propulsive performance for $B_0=600$~G.}

\section{Experimental validation}
\label{sec:valid}

\begin{table*}[!htb]
\caption{\label{tab:validparams} Validation parameters}
\lineup
\centering
\begin{indented}
\item[]\begin{tabular}{llll}
\br
Parameter &                       & Unscaled             & Scaled$^a$               \\ \mr
Reference Plasma Density                                 & $n_*$ [m$^{-3}$]      & $5\times10^{16}$     & -                    \\
Ion Mass (Ar)                                            & $m_i$ [kg]          & $6.63\times10^{-26}$ & $1.66\times10^{-27}$ \\
Reference Ion Temperature                                & $T_{i*}$ [K]       & 298                  & -                    \\
Reference Electron Temperature                           & $T_{e*}$ [eV]      & 9                    & -                    \\
Ion Speed                                           & $u_{i0}$ [ms$^{-1}$] & 3724                 & 23520      \\ \mr
Axial Domain Length & $L_z$ [m] & 0.25 & - \\
Radial Domain Length & $L_r$ [m] & 0.13 & - \\\br       
\end{tabular}
\item[] $^a$ $f=40$, $\gamma=33.1$.
\end{indented}
\end{table*}

The results of the PIC are compared against the measurements performed in the \textit{Piglet} Helicon plasma reactor, filled with argon gas at $0.04$~Pa as described in reference \cite{b:lafleur2011}. {The magnetic configuration considered is generated by an electromagnet, referred to as the \textit{source coil} in reference \cite{b:lafleur2011}, which provides a throat intensity of $B_0 = 4$~G.} The validation input parameters are given in \tref{tab:validparams}. {The domain is a truncation ($L_z=0.25$~m,~$L_r=0.13$~m) of the physical vacuum chamber used in the experiment (length $0.288$~m radius $0.16$~m), such to allow the use of the open boundaries. The plasma reference properties were taken directly from the experimentally measured values within the source tube, where $u_{i0}=0.8c_{*}$.}

{In the experiment, electron density was evaluated with a Langmuir probe, with the local plasma potential obtained from a retarding field energy analyser (RFEA).} The results along the axis of the discharge, have been reported in \fref{fig:valid_plot}(a) and \fref{fig:valid_plot}(b). The experimental plasma potential has been normalised so as that $\phi_0=0$ at $z=0$. Due to the possible overestimation of the Langmuir probe sheath area by about 15\% \cite{b:lafleur2011}, the errorbars associated to the number density are asymmetrical between -5\% and 27\%. RFEA measurements have a given uncertainty of $\pm$5\%.

Focusing on the axial density profile in \fref{fig:valid_plot}(a), the experimental trend is reproduced by the PIC model. The maximum local error is 10\% at $z=0.05$~m, well within the uncertainty bands reported. Concerning the plasma potential profile of \fref{fig:valid_plot}(b), the PIC model again repeats the experimental trend within the quoted uncertainty. However, the PIC slightly overestimates the potential in the downstream region of the plume, up to $1.6$~V at the open boundary. Critically, the potential drop calculated by the PIC is $\phi_\infty=-37.7$~V, in excellent agreement with the local valued $V_p=37.2$~V directly measured in reference \cite{b:lafleur2011}. This, as well as the similarity in potential gradient observed at $z=0.25$~m in \fref{fig:valid_plot}(b), confirms the reliability of the open boundary conditions in the new model.

\section{Physical analysis}
\label{sec:results}

{In this section, the plasma profiles are first examined. Second, the effect of varying $B_0$ on the global parameters,} including the propulsive performance indicators, is presented. A more detailed discussion is then given on the nature of electron thermodynamics in the MN, with particular focus on the role of collisions.

\subsection{Plasma Profiles}
\label{sec:profiles}
\begin{figure*}[!htb]
    \centering
    \includegraphics[width=\linewidth]{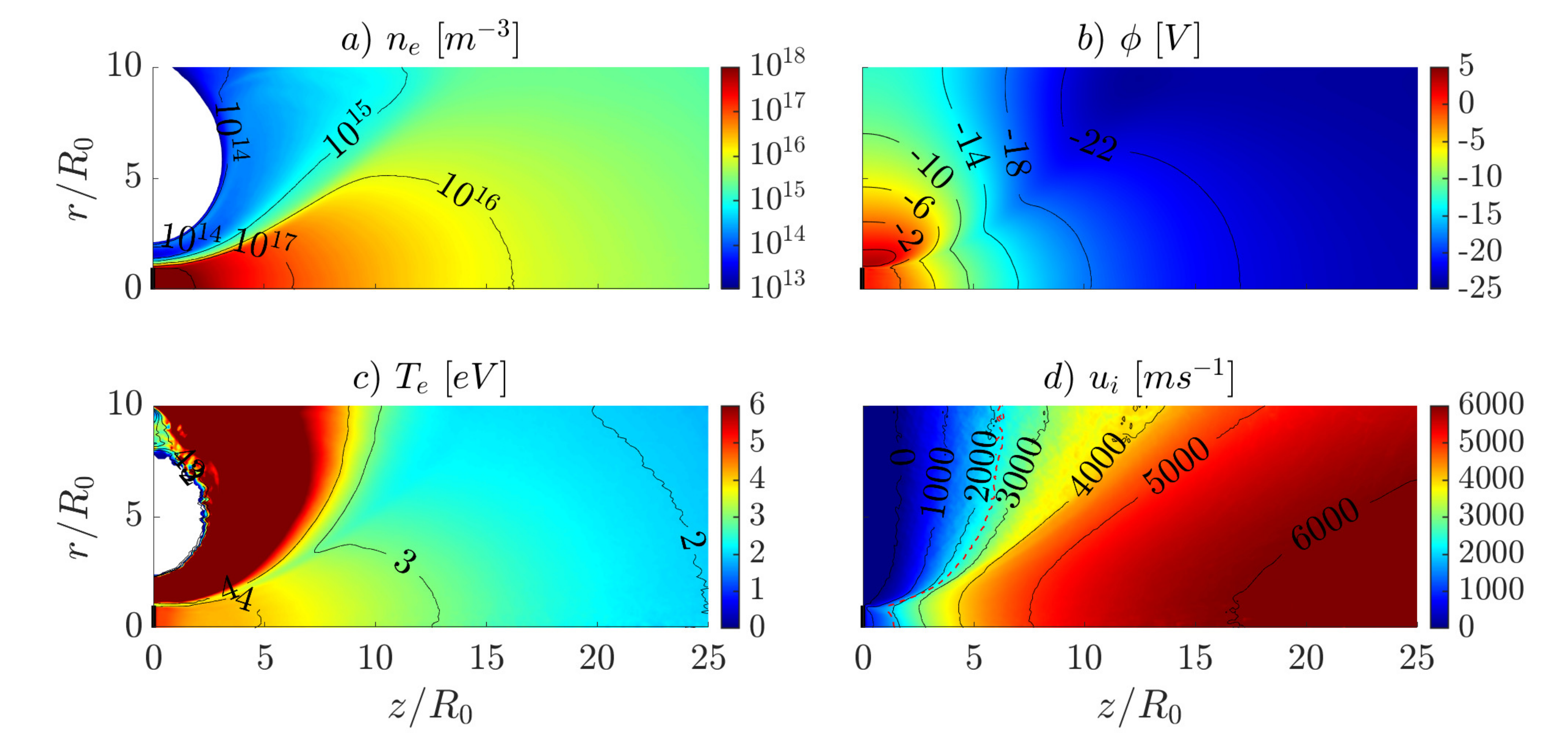}
    \caption{\centering Plasma field properties for the $B_0=600~$G case: (a) electron number density $n_e$; (b) plasma potential $\phi$; (c) electron temperature $T_e$; (d) ion axial velocity $u_i$.}
    \label{fig:spatial_plot}
\end{figure*}
\begin{figure*}[!htb]
    \centering
    \includegraphics[width=\linewidth]{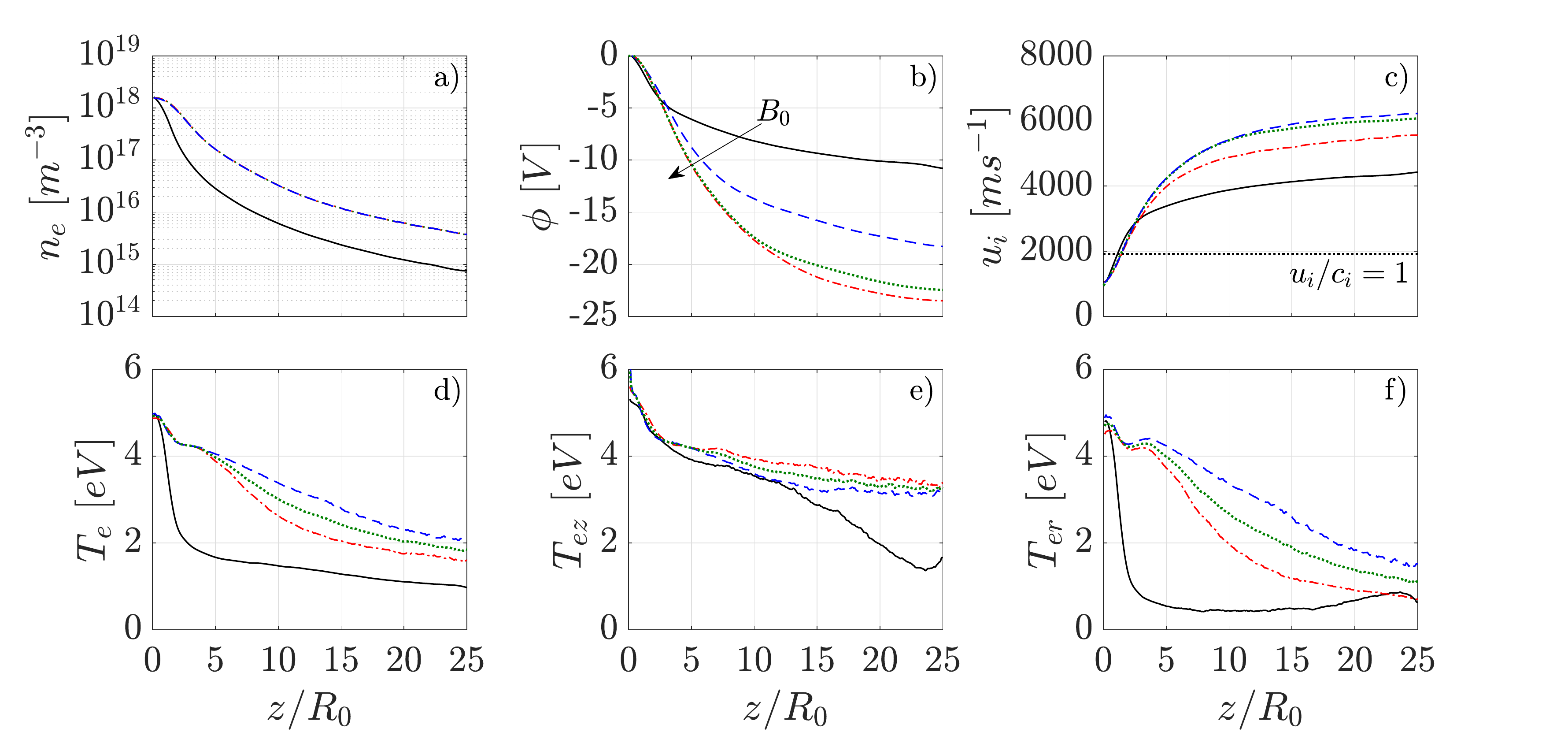}
    \caption{\centering Plasma on-axis profiles for $B_0=0$ (\full), $B_0=100$~G (\textcolor{red}{\chain}), $B_0=300$~G (\textcolor{green}{\dotted}) and $B_0=600$~G (\textcolor{blue}{\dashed}): {(a) electron number density $n_e$, (b) plasma potential $\phi$, (c) ion axial velocity $u_i$, (d) electron temperature $T_e$, (e) axial electron temperature $T_{ez}$, (f) radial electron temperature $T_{er}$.}}
    \label{fig:onaxis_plot}
\end{figure*}
\Fref{fig:spatial_plot} shows the 2D spatial fields for the $B_0=600$~G case, including (a) the electron number density, (b) plasma potential, (c) electron temperature and (d) ion axial velocity. The plasma {expansion} follows the magnetic field lines which determine the divergence of the plume. {Specifically, the plasma properties propagate monotonically downstream under the dominance of a self-consistently developed ambipolar electric field \cite{takahashi2019helicon}. Electron density drops outside the periphery of the plume, {with} an electron void occurring near the thruster outlet in \fref{fig:spatial_plot}(a).}

A notable feature is observed in the potential field of \fref{fig:spatial_plot}(b): a radially non-monotonic dependence. This is characterised by an effective potential well along the vacuum interface line (the outermost magnetic field line starting at edge of thruster outlet boundary). This has been noted in a number of magnetically confined plasmas \cite{b:collard2019,b:chen2020} and can be interpreted as the consequence of charge separation that results from ions with sufficient radial energy surpassing the attached electron fluid, causing secondary ion expansion beyond the vacuum interface line. A potential barrier forms to counter this radial ion inertia and return the ion streamlines back toward the MN-aligned electron trajectories. This is clearly seen in \fref{fig:spatial_plot}(d) by the radial discontinuity in ion axial velocity along the vacuum interface line.

{{It should be noted that the potential peak near the thruster outlet, $\sim3.5$~V, is not a usual feature observed in collisionless PIC simulations \cite{b:chen2020}. Its presence can be justified in the role played by collisions.} Radially accelerating charge-exchange (CEX) ions, combined with increased electron collisional cross-field mobility, enhances the secondary ion expansion, increasing the positive space charge and hence the strength of the potential barrier. {Moreover, the large ion mass (the propellant gas is xenon) is expected to enhance the amplitude of the potential peak, since} heavier ions should require a larger electric field to return their trajectories into the MN.}

Cooling of electrons downstream in \fref{fig:spatial_plot}(c) occurs as electron thermal energy is evidently converted to ion kinetic energy, facilitated by the ambipolar potential drop, with the magnetic field acting as a mediating factor. The region of high $T_e>6$~eV at the radial peripheries of the thruster outlet (beyond the vacuum interface line) occurs since only the most highly-energetic electrons can detach early from the MN.

In order to provide a more quantitative comparison on the effect of the MN field strength, 1D plasma profiles have been sampled along the z-axis in \fref{fig:onaxis_plot} for $B_0=0,~100,~300$, and $600$~G. From \fref{fig:onaxis_plot}(a), {the application of the MN yields higher plasma density because of the increased radial confinement of the plume,} but there is no significant change between $100$ and $600$~G. The MN effect tends to increase electron current due to the $\mathbf{v}_e\times \mathbf{B}$ force exerted on electrons. As a consequence, the potential drop increases with $B_0$ (\fref{fig:onaxis_plot}(b)), so as to maintain the current-free plume. Consistently, the ion axial speed increases (\fref{fig:onaxis_plot}(c)). It is interesting to note that the potential drop/acceleration occurs in a larger axial span in the magnetised cases. This might be explained with mass conservation, since the ion beam divergence for the unmagnetised case is much higher, so a faster geometric expansion is expected. 

The electron cooling is reduced with the increase in $B_0$ (\fref{fig:onaxis_plot}(d)). It is reasonable to associate the slower cooling of the electrons to the increased plume confinement and so a reduced loss of energy through the periphery of the MN; this enables more electron energy to be available downstream. Strong temperature anisotropy is developed as shown in \fref{fig:onaxis_plot}(e) and (f). {Electron temperature along the z-axis ($T_{ez}$) decays to a non-zero value ($\sim3.6$~eV regardless the MN strength), while $T_{er}$ decays to near-zero. Interestingly,
no magnetic field means a higher divergence of the plume, and this decreases greatly $T_{er}$ in the region close to the thruster outlet, resulting in the smaller mean temperature seen in \fref{fig:onaxis_plot}(d). On the other hand, the anisotropy on $T_e$ increases downstream for both MN strengths which confirms the conversion of electron internal energy into ion axial kinetic energy \cite{b:martinez2015}.}

{Finally, a deeper analysis is conducted to explain why, {independent of} the MN strength, the plasma potential and axial ion speed are almost equal for $z\lesssim 3R_0$.} {The relatively high neutral density near the thruster outlet leads to CEX ion collisions, which act as a drag term on the ions. To assess this, the CEX mean free path $\lambda_{CEX}$ can be compared to the characteristic length scale of axial ambipolar acceleration $\lambda_{\nabla\phi}=|(\phi-\phi_\infty)/E_z|$ \cite{b:griffiths2013}. $\lambda_{CEX}$ can be estimated from the PIC simulation as $\lambda_{CEX} \sim |\mathbf{v}_i|/\nu_{CEX}$, where $\nu_{CEX}$ is the CEX collision frequency taken from the MCC module. If the CEX mean free path is shorter than the acceleration length scale, the ions are experiencing a drag force through frequent CEX collisions. Otherwise, the ion acceleration is not significantly impeded by these collisions.
\begin{figure}[!tb]
    \centering
    \includegraphics[width=\linewidth]{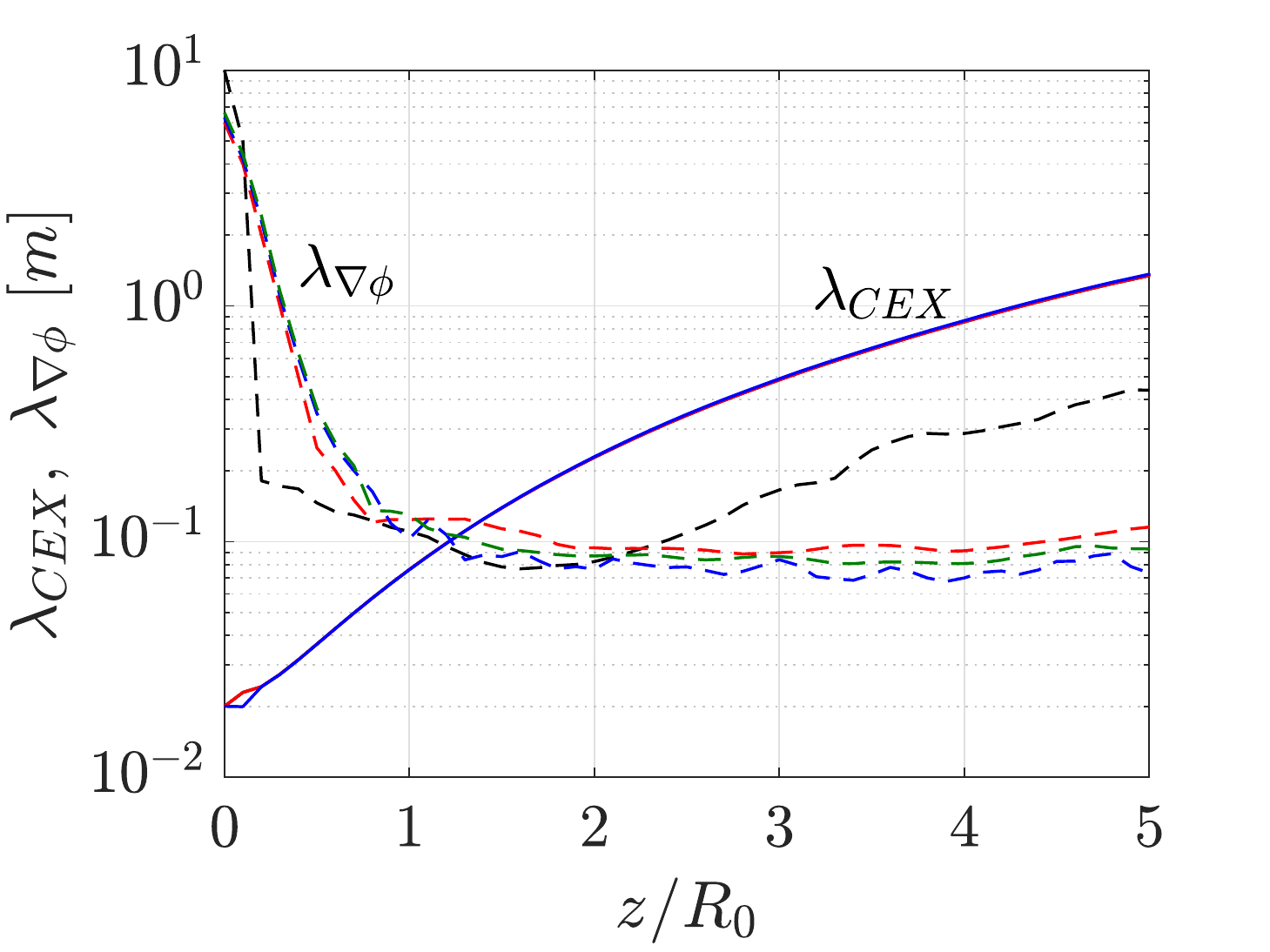}
    \caption{\centering Comparison of the on-axis CEX mean free path (solid lines) to the ion ambipolar acceleration length (dashed lines) for $B_0=0$ (\full), $100$~G (\textcolor{red}{\full}), $300$~G (\textcolor{green}{\full}) and $600$~G (\textcolor{blue}{\full}).}
    \label{fig:mfp}
\end{figure}
The on-axis length scales are given in {\fref{fig:mfp}}. The axial location at which the CEX mean free path becomes equal to the acceleration length scale occurs at $z\sim1.2R_0$ independently of the MN field strength. The action of this CEX drag may explain why the potential drop, and subsequent ion velocity, in the near thruster outlet region is identical in each case presented in \fref{fig:onaxis_plot}. This behaviour is unlikely to hold true if the plasma source was included in the model; the effect of the MN field strength would alter the ionisation efficiency, and therefore the ratio of plasma to neutral density.} {Drag from CEX collisions also explains why the plasma choke point ($u_{iz}=c_i$) is located downstream from the MN throat at $z\sim 1.5R_0$, a consequence also observed in reference \cite{b:collard2019}.}
%, as illustrated by the sonic line in Figure \ref{fig:spatial_plot}(d).} 

%
%
\begin{figure*}[!htb]
    \centering
    \includegraphics[width=\linewidth]{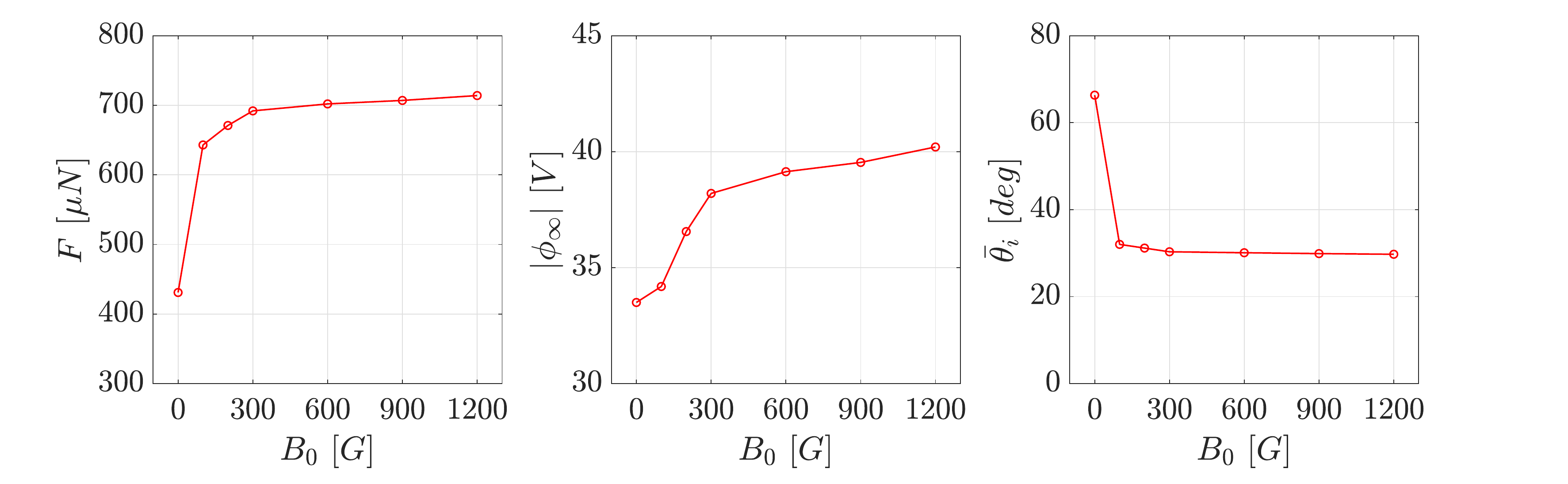}
    \caption{\centering Effect of MN field strength on (a) Thrust $F$; (b) potential drop $|\phi_\infty|$ and (c) Ion divergence angle $\bar{\theta}_{i}$}
    \label{fig:param}
\end{figure*}
\subsection{Global Parameters}
\label{sec:global}

The {propulsive} performance of the MNs is illustrated in \fref{fig:param}. As the magnetic field strength increases, thrust also increases at a diminishing rate. {The magnetic thrust in the $100$~G case accounts for approximately 50\% of the total. At $1200$~G the magnetic thrust is around 65\%,} showing there is limit to the performance enhancement the MN can provide. The gain in thrust between the unmagnetised and magnetised cases is namely due to the increased radial confinement of ions and the corresponding increase in downstream density. Indeed there is a clear trend in the thrust of \fref{fig:param}(a) and the plume divergence $\bar{\theta}_i=\langle cos^{-1}(\mathbf{v}_i\cdot\hat{\mathbf{z}}/|\mathbf{v}_i|) \rangle_i$ of \fref{fig:param}(c). As $B_0$ is further increased, the additional thrust is due to the higher ion axial velocity achieved in association with the greater potential drop, seen in \fref{fig:param}(b). The potential drop must increase to balance a growing electron current induced by the MN effect. Note, that in practice, the influence of the MN on performance is far more complex because it also affects the source region. The magnetic field strength affects the source confinement \cite{b:magarotto2020}, deposition of power into the plasma \cite{takahashi2019helicon}, and the electron distribution function \cite{b:lafleur2015} of the discharge into the MN throat.

\subsection{Electron thermodynamics}
\label{sec:thermo}
\begin{figure}[!tb]
    \centering
    \includegraphics[width=\linewidth]{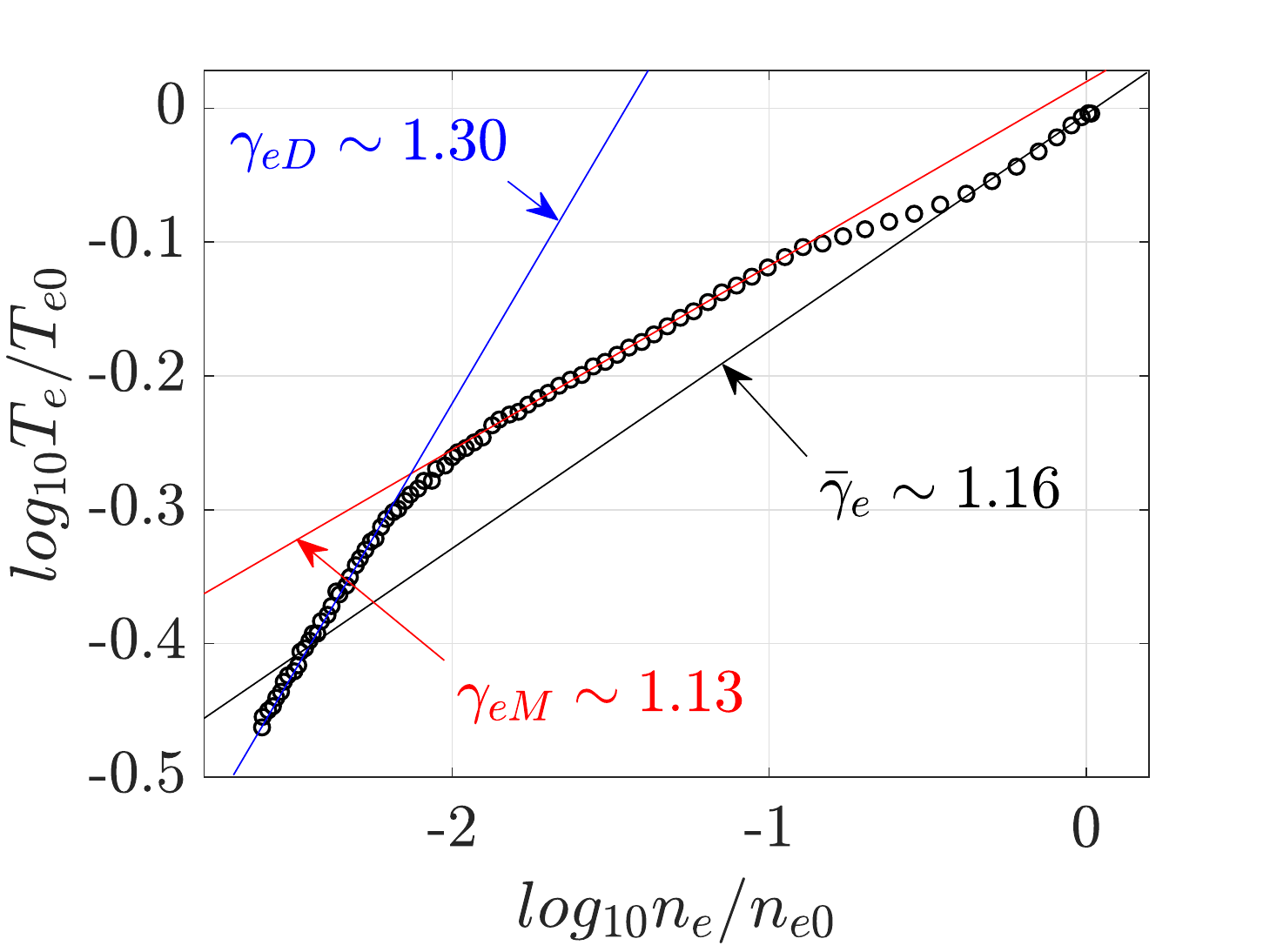}
    \caption{\centering Polytropic index fitting of the $600$~G case.}
    \label{fig:gamma_fit}
\end{figure}

The electron cooling in MN expansions may be described by a polytropic relation $T_e/T_{e0}=(n_e/n_{e0})^{\gamma_e-1}$ where $\gamma_e$ is the polytropic index \cite{b:ahedo2010}. {It can be calculated considering that 
\begin{equation}
    \gamma_e = 1+\frac{n_e}{T_e}\frac{dT_e}{dn_e}
\end{equation}
with $\gamma_e$ derived by the linear} regression of the $log_{10}T_e$ versus $log_{10}n_e$ relation, an example of which is given in \fref{fig:gamma_fit} for the $600$~G case. The average value is shown to be $\bar{\gamma}_e\sim1.16$, but it is clear that a single polytropic index cannot represent the electron cooling in the MN. Three separate regions are therefore identified for a piecewise polytropic relationship. {Nearest the thruster outlet there is a region with a mildly-adiabatic value. Further downstream, and for most of the expansion, there is a region with a near-isothermal $\gamma_{eM}\sim1.13$. Finally, farther downstream, there exists a region with a markedly greater value of $\gamma_{eD}\sim1.30$. Piecewise polytropic behaviour has also been observed in reference \cite{b:correyero2019}, where the measurements of an ECR thruster plume agree with the first and second regions identified here.}

The break between the second and third polytropic regions agrees with the locations of the MN detachment planes identified in \fref{fig:axial_thrust}. It can therefore be inferred that the second region is where the electrons are well-magnetised and frozen to the magnetic field lines. The third region represents that where the plasma has detached from the MN and so the electron cooling tends to the same as that for an unmagnetised expansion. The first region possesses an average value that lies between the indices found in the other two regions, therefore it may be assumed that here electrons are not fully attached to the MN. 
\begin{figure}[!htb]
    \centering
    \includegraphics[width=\linewidth]{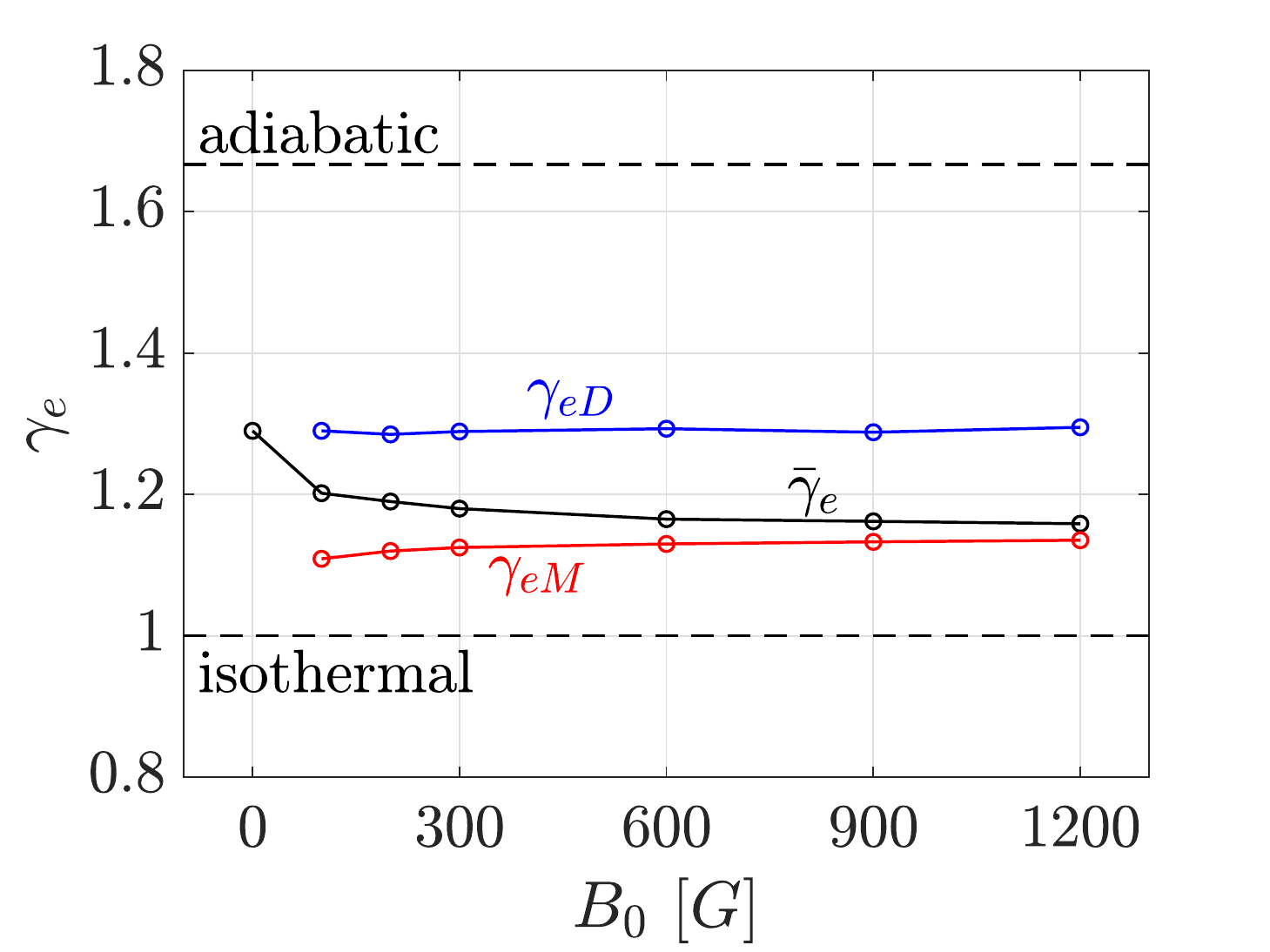}
    \caption{\centering The values of the polytropic index, taken using the linear regression, as a function of $B_0$.}
    \label{fig:gamma_b0}
\end{figure}

\Fref{fig:gamma_b0} shows the effect of the MN strength on the indices describing the different polytropic regions. Notably, the value of $\gamma_{eD}$ is approximately constant, and roughly equal to the unmagnetised average $\bar{\gamma}_e=1.29$, for all cases. This confirms this region is post plasma detachment. 
{The value of $\gamma_{eM}$ is almost constant, consistent with the mild effect that the intensity of the MN has on $n_e$ and $T_e$ profiles upstream of the detachment plane for $B_0\geq100$~G (see \sref{sec:profiles}).}
Increasing the magnetic field strength reduces the average electron cooling rate, tending to $\bar{\gamma}_e\sim 1.16$ for $600-1200$~G. {This is due to the larger axial length in which electrons are magnetised, since the detachment plane moves downstream with increasing $B_0$ (see \sref{sec:domain}).} The values found for $\bar{\gamma}_e$ are in good agreement with experiments on xenon MNs, which have reported magnitudes between 1.1 and 1.23 \cite{b:collard2019,b:correyero2019}. Measurements of $\bar{\gamma}_e$ = 1.15 $\pm$ 0.02 have been reported also in reference \cite{b:little2016} and a theoretical limit of $\bar{\gamma}_e=1.16$ was predicted in reference \cite{b:littlephd}. 

\begin{figure}[!htb]
    \centering
    \includegraphics[width=.98\linewidth]{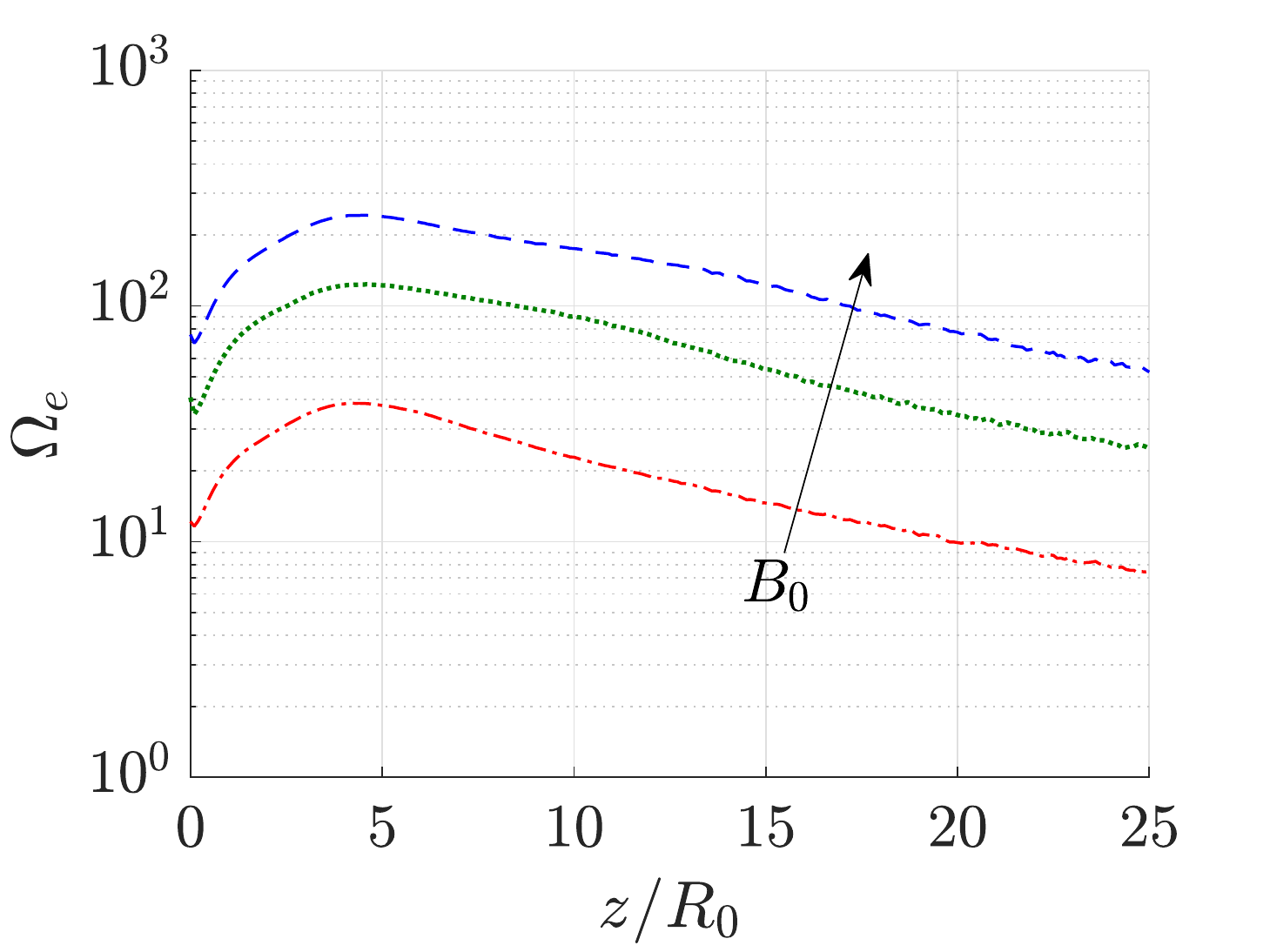}
    \caption{\centering Electron Hall parameter $\Omega_e$ along the axis for the $100$~G (\textcolor{red}{\chain}), $300$~G (\textcolor{green}{\dotted}) and $600$~G (\textcolor{blue}{\dashed}) case.}
    \label{fig:hall}
\end{figure}

{In order to interpret the behaviour of $\gamma_e$ in the region near the thruster outlet, consider the electron Hall parameter}
\begin{equation}
    \Omega_e = \frac{e|\mathbf{B}|}{m_e\nu}
\end{equation}
where $\nu=\nu_{en}+\nu_{ei}$ is the total of the electron-neutral and electron-ion collision frequencies calculated from the MCC module. {$\Omega_e$ is equal to the ratio between the cyclotron frequency and the collision frequency, so it is an indicator of the level of electron magnetisation/attachment. In \fref{fig:hall} the electron Hall parameter is shown for $100$, $300$ and $600$~G. Sufficiently upstream near the thruster outlet,} collisions act to de-magnetise the electrons. The plasma remains dominated by collisions in the near-field plume. This collisionality reduces the plasma conduction and thermalises the electron distribution, increasing the near-throat polytropic index towards the adiabatic. This precludes the accelerating action of the magnetic field until the neutral gas becomes sufficiently sparse {that the electrons become strongly magnetised again.}

\section{Conclusions}
\label{sec:conclude}

In this article, a new set of self-consistent, open boundary conditions for electron kinetics and the Poisson's equation have been introduced. They were developed to perform steady-state fully kinetic PIC simulations of plasma expansion in the MN. The newly developed boundary conditions correct the non-physical loss of electrons by reflecting them at the open boundaries according to {an energy-based criterion.} A virtual capacitor allows equal ion and electron currents to infinity at steady-state. The electric field at the external boundaries is also self-consistent with the potential drop according to a new Robin-type condition on the Poisson's equation. This approach generated a stable, steady-state plume using axisymmetric two-dimensional fully kinetic simulations for typical {operative conditions of a low-power (50~W) MEPT.}

The robustness of the new model was established by changing the value of the virtual capacitance and thus the transient evolution of the plasma potential and plume. It was demonstrated that identical steady-state solutions are obtained, and thus the boundary conditions yield results independent of the choice of capacitance. Domain-independence studies were performed to study the sensitivity of the computed number density and plasma potential to changes in the location of the domain boundaries. {The results of the new set of boundary conditions were benchmarked against experiments providing electron density and plasma potential profiles that fall within the uncertainty band of the measures.} 

{The validated model was exploited to investigate the plasma expansion in a MN. A potential well forms at the periphery of the plume to counterbalance the cross-field diffusion of the weakly-magnetized ions. The performance indicators increase with the strength of the MN because of the enhanced radial confinement and ion acceleration. Nonetheless the increase in the performance a MN can provide is limited at about 70\% \cite{b:martinez2015}. The electron cooling is affected by collisions.}  
\begin{itemize}
    \item There exists a three-region polytropic cooling regime, defined by a partially-detached near-exit region, a strongly attached near-isothermal region, and a more adiabatic detached region. 
    {\item The third region occurs downstream the detachment plane, here the polytropic index is equal to the one of an unmagnetised plume.
    \item The average polytropic index decreases with the strength of the MN since the detachment plane moves downstream and so the electrons are strongly attached to the MN in a larger portion of space. An inferior limit of $\bar{\gamma}_e=1.16$ was found.
    \item The electron cooling in the near-exit region is determined by collisions that tend to partially-detach electrons from the MN.}
\end{itemize}
{Moreover, the CEX collisions act as a drag term on the ions in the near-exit region, such that ion acceleration matches the unmagnetised one.}

The new boundary conditions offer a valuable tool in the performance evaluation and optimisation of MEPTs (and indeed unmagnetised thrusters also), reducing the computational time compared to the large domains required by other models {\cite{b:brieda2018}}. Future work will involve iterative coupling of the PIC to the fluid model \textit{3D-VIRTUS}, developed to simulate the source region of such devices \cite{b:iac2021,b:magarotto2020_2,b:souhairAIP,b:souhairPP}. The boundary conditions will also be applied within the framework of a 3D PIC code \cite{b:difede2021} to assess the limitations of the 2D axisymmetric assumptions, and analyse the plume interactions with non-axisymmetric spacecraft bodies.

\ack 
This work was partially funded by Technology for Innovation and Propulsion (T4i) S.p.A.. S. Andrews was also supported by a scholarship from the European Union Horizons 2020 MSCA-RISE project PATH, under grant agreement 734629.

\appendix
\section*{Appendix. Derivation of the Robin boundary condition}
\label{sec:appendix}
\setcounter{section}{1}
{Generally, the electric potential generated by $N_p$ charged particles at positions $\mathbf{r}_p$ is obtained by the sum of the individual point charges.} Consider the potential very far from the localised charge distribution of the plume; it resembles a total point-like charge. Griffiths \cite{b:griffiths2013} provides a multipole expansion for the approximate potentials at large distances $\mathbf{r}$ from a localised charge distribution, which is modified here to the discrete form and referenced to the potential at infinity,
\begin{eqnarray}
& \phi(\mathbf{r}) = \frac{1}{4\pi\epsilon_0}\sum_{s=0}^\infty\frac{1}{|\mathbf{r}|^{s+1}}\sum^{N_p}_{p=1} (\mathbf{r}_p)^s {P}_s(\cos\theta_p)q_p + \phi_\infty
\nonumber \\ &~
\end{eqnarray}
{where ${P}_s$ is the Legendre polynomial operator,} and $\theta_p$ is the angle between $\mathbf{r}$ and $\mathbf{r}_p$. This is the multipole expansion of $\phi$ in powers of $1/|\mathbf{r}|$. The first term ($s=0$) is the monopole contribution $1/|\mathbf{r}|$; the second ($s=1$) is the dipole contribution $1/|\mathbf{r}|^2$; and so on. Eq.~A.1 is exact, but it is useful primarily as an approximation scheme. The lowest non-zero term in the expansion provides the approximate potential at large $\mathbf{r}$. At very large $\mathbf{r}$, the expansion is dominated by the monopole term. So the potential far from the thruster outlet is, to good approximation,  
\begin{equation}
\phi(\mathbf{r}) = \frac{1}{4\pi\epsilon_0}\frac{1}{|\mathbf{r}|}\sum^{N_p}_{p=1}q_p + \phi_\infty
\end{equation}
where $\sum^{N_p}_{p=1}q_i$ is the total net charge in the plume, which, for a partially confined plasma, is not null {(i.e., plasma is non-neutral in the sheath formed by the walls)}. Eq.~A.2 is the standard $1/r$ monopole decay into vacuum. {Taking the gradient of the potential in Eq.~A.2 gives,
\begin{equation}
\nabla\phi(\mathbf{r}) = -\frac{1}{4\pi\epsilon_0}\frac{\mathbf{r}}{|\mathbf{r}|^3}\sum^{N_p}_{p=1}q_p
\end{equation}
A formal relation between $\phi$ and $\nabla\phi$ is then obtained by substituting Eq.~A.2 into Eq.~A.3,
\begin{equation}
-\nabla\phi(\mathbf{r}) = \frac{\mathbf{r}}{|\mathbf{r}|^2}\left(\phi(\mathbf{r})-\phi_\infty\right)
\end{equation}
The projection of Eq.~A.4 along a unit vector $\hat{\mathbf{n}}$ subsequently reads,}
\begin{equation}
  \frac{\partial\phi}{\partial\hat{\mathbf{n}}}+\frac{\hat{\mathbf{n}}\cdot{\mathbf{r}}}{\mathbf{r}\cdot\mathbf{r}}\left(\phi(\mathbf{r})-\phi_\infty\right) = 0
\end{equation}
which, when $\hat{\mathbf{n}}$ is the normal to the domain boundary, provides an open boundary condition on the Poisson's equation for plasma expansion into vacuum.
\section*{References}
\bibliographystyle{my-iopart-num}
\bibliography{bibliography}

\providecommand{\newblock}{}
\providecommand{\url}[1]{{\tt #1}}
\providecommand{\urlprefix}{}
\providecommand{\href}[2]{#2}
\begin{thebibliography}{10}

\bibitem{b:keidar2014}
Keidar M, Zhuang T, Shashurin A, Teel G, Chiu D, Lukas J, Haque S and Brieda L
  2014 \href{https://doi.org/10.1088/0741-3335/57/1/014005}{{\em Plasma Physics
  and Controlled Fusion\/} {\bf 57}}

\bibitem{takahashi2019helicon}
Takahashi K 2019 {\em Reviews of Modern Plasma Physics\/} {\bf 3} 1--61

\bibitem{b:boswell2003}
Boswell R~W and Charles C 2003 The helicon double layer thruster {\em 28$^{th}$
  International Electric Propulsion Conference\/} IEPC-2003-332 (Toulouse,
  France)

\bibitem{b:shinohara2014}
Shinohara S, Nishida H, Tanikawa T, Hada T, Funaki I and Shamrai K~P 2014 {\em
  IEEE Transactions on Plasma Science\/} {\bf 42} 1245--54

\bibitem{b:merino2015}
Merino M {\em et~al.\/} 2015 Design and development of a 1 kw-class helicon
  antenna thruster {\em 34$^{th}$ International Electric Propulsion
  Conference\/} IEPC-2015-297 (Kobe, J)

\bibitem{b:cannat2015}
Cannat F, Lafleur T, Jarrige J, Chabert P, Elias P~Q and Packan D 2015 {\em
  Physics of Plasmas\/} {\bf 22} 053503

\bibitem{b:boxberger2019}
Boxberger A, Behnke A and Herdrich G 2019 Current advances in optimization of
  operative regimes of steady state applied field mpd thrusters {\em 36$^{th}$
  International Electric Propulsion Conference\/} IEPC-2015-585 (Vienna,
  Austria)

\bibitem{b:merino2016}
Merino M and Ahedo E 2016 Space plasma thrusters: Magnetic nozzles for {\em
  Encyclopedia of Plasma Technology\/} vol~2 ed Shohet J (New York: Taylor \&
  Francis) p 1329–1351 1st ed

\bibitem{b:ahedo2010}
Ahedo E and Merino M 2010 {\em Physics of Plasmas\/} {\bf 17} 073501

\bibitem{b:chen2015}
Chen F~F 2015 {\em Plasma Sources Science and Technology\/} {\bf 24} 014001

\bibitem{b:magarotto2019dep}
Magarotto M, Melazzi D and Pavarin D 2019 {\em Journal of Plasma Physics\/}
  {\bf 85} 905850404

\bibitem{b:manente2019}
Manente M, Trezzolani F, Magarotto M, Fantino E, Selmo A, Bellomo N, Toson E
  and Pavarin D 2019 {\em Acta Astronautica\/} {\bf 157} 241--9

\bibitem{b:bellomo2021}
Bellomo N, Magarotto M, Manente M {\em et~al.\/} 2021 {\em CEAS Space
  Journal\/}  1868--2510

\bibitem{b:magarotto2020}
Magarotto M, Manente M, Trezzolani F and Pavarin D 2020 {\em IEEE Transactions
  on Plasma Science\/} {\bf 48} 835--44

\bibitem{b:magarotto2020_2}
Magarotto M, Melazzi D and Pavarin D 2020 {\em Computer Physics
  Communications\/} {\bf 247} 106953

\bibitem{b:magarotto2020_3}
Magarotto M and Pavarin D 2020 {\em IEEE Transactions on Plasma Science\/} {\bf
  48} 2723--35

\bibitem{b:longmier2011}
Longmier B~W, Bering E~A, Carter M~D, Cassady L~D, Chancery W~J, D{\'{\i}}az
  F~R~C, Glover T~W, Hershkowitz N, Ilin A~V, McCaskill G~E, Olsen C~S and
  Squire J~P 2011 \href{https://doi.org/10.1088/0963-0252/20/1/015007}{{\em
  Plasma Sources Science and Technology\/} {\bf 20} 015007}

\bibitem{b:olsen2015}
{Olsen} C~S, {Ballenger} M~G, {Carter} M~D, {Díaz} F~R~C, {Giambusso} M,
  {Glover} T~W, {Ilin} A~V, {Squire} J~P, {Longmier} B~W, {Bering} E~A and
  {Cloutier} P~A 2015 \href{https://doi.org/10.1109/TPS.2014.2321257}{{\em IEEE
  Transactions on Plasma Science\/} {\bf 43} 252--68}

\bibitem{b:lafleur2015}
Lafleur T, Cannat F, Jarrige J, Elias P and Packan D 2015 {\em Plasma Sources
  Science and Technology\/} {\bf 24} 065013

\bibitem{b:kaganovich2020}
Kaganovich I, Smolyakov A, Raitses Y, Ahedo E, Mikellides I, Jorns B, Taccogna
  F, Gueroult R, Tsikata S, Bourdon A, Boeuf j~p, Keidar M, Powis A, Merino M,
  Cappelli M, Hara K, Carlsson J, Fisch N, Chabert P and Fruchtman A 2020
  \href{https://doi.org/10.1063/5.0010135}{{\em Physics of Plasmas\/} {\bf 27}
  120601}

\bibitem{b:merino2018kinetic}
Merino M, Maurino J and Ahedo E 2018 {\em Plasma Sources Science and
  Technology\/} {\bf 27}(03) 035013

\bibitem{b:sasoh1994}
Sasoh A 1994 \href{https://doi.org/10.1063/1.870847}{{\em Physics of Plasmas\/}
  {\bf 1} 464--9}

\bibitem{b:martinez2015}
Martinez-Sanchez M, Navarro-Cavall{\'e} J and Ahedo E 2015 {\em Physics of
  Plasmas\/} {\bf 22} 053501

\bibitem{b:ahedo2020}
Ahedo E, Correyero S, Navarro-Cavall{\'{e}} J and Merino M 2020
  \href{https://doi.org/10.1088/1361-6595/ab7855}{{\em Plasma Sources Science
  and Technology\/} {\bf 29} 045017}

\bibitem{b:kim2005}
Kim H, Iza F, Yang S, Radmilovi{\'c}-Radjenovi{\'c} M and Lee J 2005 {\em
  Journal of Physics D: Applied Physics\/} {\bf 38}(19) R283--301

\bibitem{b:difede2021}
Di~Fede S, Magarotto M, Andrews S and Pavarin D 2021
  \href{https://doi.org/10.1017\/S0022377821001057}{{\em Journal of Plasma
  Physics\/} }

\bibitem{b:iac2021}
Magarotto M, Di~Fede S, Souhair N, Andrews S, Manente M, Ponti F and Pavarin D
  2021 Numerical suite for magnetically enhanced plasma thrusters {\em
  72$^{nd}$ International Astronautical Congress\/} IAC-21 C4.6.3 (Dubai, UAE)

\bibitem{b:porto2019}
Porto J and Elias P~Q 2019 Full-pic simulation of an {ECR} plasma thruster with
  magneticnozzle {\em 36$^{th}$ International Electric Propulsion Conference\/}
  IEPC-2019-232 (Vienna, Austria)

\bibitem{b:bipic}
Gallina G, Magarotto M, Manente M and Pavarin D 2019 {\em Journal of Plasma
  Physics\/} {\bf 85} 905850205

\bibitem{b:alvaro2021}
Sanchez-Villar A, Zhou J, Ahedo E and Merino M 2021 {\em Plasma Sources Science
  and Technology\/} {\bf 30} 045005

\bibitem{b:birdsall1991}
{Birdsall} C~K 1991 {\em IEEE Transactions on Plasma Science\/} {\bf 19} 65--85

\bibitem{b:li2019}
Li M, Merino M, Ahedo E and Tang H 2019 {\em Plasma Sources Science and
  Technology\/} {\bf 28} 034004

\bibitem{b:merino2019iepc}
Nuez J, Merino M and Ahedo E 2019 Fluid-kinetic propulsive magnetic nozzle
  model in the fully magnetized limit {\em 36$^{th}$ International Electric
  Propulsion Conference\/} IEPC-2019-254 (Vienna, Austria)

\bibitem{b:brieda2018}
Brieda L 2018 {\em IEEE Trans. Plasma Sci.\/} {\bf 46} 556--62

\bibitem{b:briedaThesis}
Brieda L 2005 {\em Development of the DRACO ES-PIC code and Fully-Kinetic
  Simulation of Ion Beam Neutralization\/} Ph.D. thesis Virginia Polytechnic
  Institute

\bibitem{b:levin2018}
Jambunathan R and Levin D 2018 {\em J. Comput. Phys.\/} {\bf 373} 571--604

\bibitem{b:hu2017}
Hu Y and Wang J 2017 {\em Phys. Plasmas\/} {\bf 24} 033510

\bibitem{b:levin2020}
{Jambunathan} R and {Levin} D~A 2020
  \href{https://doi.org/10.1109/TPS.2020.2968887}{{\em IEEE Transactions on
  Plasma Science\/} {\bf 48} 610--30}

\bibitem{b:chen2020}
Chen Z, Wang Y, Tang H~B, Ren J, Li M, Zhe Z, Cao S and Cao J 2020
  \href{https://doi.org/10.1103/PhysRevE.101.053208}{{\em Physical Review E\/}
  {\bf 101} 053208}

\bibitem{b:lafleur2011}
Lafleur T, Charles C and Boswell R 2010 {\em Physics of Plasmas\/} {\bf 17}
  043505

\bibitem{b:brieda2012}
Brieda L and Keidar M 2012 Development of the starfish plasma simulation code
  and update on multiscale modeling of hall thrusters {\em 48$^{th}$
  AIAA/ASME/SAE/ASEE Joint Propulsion Conference \& Exhibit\/} AIAA 2012-4015
  (Atlanta, GA, USA)

\bibitem{b:iac2019}
Andrews S and Berthoud L 2019 Effect of ion thruster
  plume-thermosphere/ionosphere interaction on satellite drag in very low earth
  orbit {\em 70$^{th}$ International Astronautical Congress\/} IAC-19 C4.5.1
  (Washington, DC, USA)

\bibitem{b:zolotukhin2020}
Zolotukhin D~B, Daniels K~P, Brieda L and Keidar M 2020
  \href{https://doi.org/10.1103/PhysRevE.102.021203}{{\em Phys. Rev. E\/} {\bf
  102}(2) 021203}

\bibitem{b:birdsall2005}
Birdsall C~K and Langdon A~B 2005 {\em Plasma physics via computer
  simulation\/} (New York NY, USA: Taylor \& Francis Group)

\bibitem{b:ruyten1993}
Ruyten W 1993 {\em Journal of Computational Physics\/} {\bf 105} 224--32

\bibitem{b:szabo}
Szabo J 2001 {\em Fully kinetic numerical modeling of a plasma thruster\/}
  Ph.D. thesis Massachusetts Institute of Technology

\bibitem{b:bird1998}
Bird G~A 1994 {\em Molecular gas dynamics and the direct simulation of gas
  flows\/} (Clarendon: Oxford University press)

\bibitem{LXCat}
Pancheshnyi S, Biagi S, Bordage M, Hagelaar G, Morgan W, Phelps A and Pitchford
  L 2012 {\em Chem. Phys.\/}  148--53

\bibitem{b:griffiths2013}
Griffiths D~J 2013 {\em Introduction To Electrodynamics\/} (Boston, USA:
  Pearson)

\bibitem{chen1999upper}
Chen F~F and Blackwell D~D 1999 {\em Physical review letters\/} {\bf 82}(13)
  2677--80

\bibitem{pottinger2011performance}
Pottinger S, Lappas V, Charles C and Boswell R 2011 {\em Journal of Physics D:
  Applied Physics\/} {\bf 44} 235201

\bibitem{b:collard2019}
Collard T and B J 2019 {\em Plasma Sources Science and Technology\/} {\bf 28}
  105019

\bibitem{b:correyero2019}
Correyero S, Jarrige J, Packan D and Ahedo E 2019 {\em Plasma Sources Science
  and Technology\/} {\bf 28} 095004

\bibitem{b:little2016}
Little J~M and Choueiri E~Y 2016 {\em Phys. Rev. Lett.\/} {\bf 117}(22) 225003

\bibitem{b:littlephd}
Little J 2015 {\em Performance scaling of magnetic nozzles for electric
  propulsion\/} Ph.D. thesis Princeton University

\bibitem{b:souhairAIP}
Souhair N, Magarotto M, Ponti F and Pavarin D 2021
  \href{https://doi.org/10.1063\/5.0066221}{{\em AIP Advances\/} {\bf 11}(11)
  115016}

\bibitem{b:souhairPP}
Souhair N, Magarotto M, Majorana E, Ponti F and Pavarin D 2021
  \href{https://doi.org/10.1063\/5.0057494}{{\em Physics of Plasmas\/} {\bf
  28}(9) 093504}

\end{thebibliography}

\end{document}